\documentclass[twocolumn,showpacs,preprintnumbers,amsmath,amssymb,aps,prb]{revtex4}

\usepackage{graphicx}
\usepackage{dcolumn}
\usepackage{bm}

\usepackage{color}
\usepackage{soul} 

\usepackage{appendix}

\newcommand{\figwidth}{0.75\linewidth}
\newcommand{\figwidthb}{0.85\linewidth}

\graphicspath{{./figs/}}
\DeclareGraphicsExtensions{.pdf}

\begin{document}

\title{Mesoscopic behavior of the transmission phase through confined correlated electronic systems}

\author{Rafael A.\ Molina}

\affiliation{Instituto de Estructura de la Materia, CSIC, Serrano 123, 28006 Madrid, Spain.}

\author{Peter Schmitteckert}

\affiliation{Institute of Nanotechnology, Karlsruhe Institute of Technology, 76344 Eggenstein-Leopoldshafen, Germany \\
Center of Functional Nanostructures, Karlsruhe Institute of Technology, 76131 Karlsruhe, Germany}

\author{Dietmar Weinmann}

\author{Rodolfo A.\ Jalabert}

\affiliation{Institut de Physique et Chimie des Mat{\'e}riaux de
Strasbourg, UMR 7504, CNRS-UdS, \\
23 rue du Loess, BP 43, 67034 Strasbourg Cedex 2, France}

\author{Philippe Jacquod}

\affiliation{Physics Department, University of Arizona
1118 E.\ Fourth Street, P.O. Box 210081, Tucson, AZ 85721, USA \\
D\'epartement de Physique Th\'eorique, Universit\'e de Gen\`eve\\
24 Quai Ernest Ansermet, 1211 Gen\`eve, Switzerland }

\begin{abstract}

We investigate the effect of electronic correlations on the transmission phase of quantum 
coherent scatterers, 
considering quantum dots in the Coulomb blockade regime connected to two single-channel leads.
We focus on transmission zeros and the associated $\pi$-phase lapses that have been observed in 
interferometric experiments. We numerically explore two types of models for quantum dots:
(i) lattice models with up to eight sites, and (ii) resonant level models with up to six levels.
We identify different regimes of parameters where the presence of electronic correlations is responsible 
for the increase or the decrease of the number of transmission zeros \textit{vs.} electrochemical 
potential on the dot. We show that within the two models considered, interaction effects do not reproduce 
the universal behavior of alternating resonances and phase lapses, experimentally observed in 
many-electron Coulomb blockaded dots. 
\end{abstract}


\maketitle

\section{Introduction}
\label{Sec:Introduction}

Quantum coherent effects in electronic transport such as Aharonov-Bohm (AB) conductance oscillations, 
weak localization, and universal conductance fluctuations~\cite{Imry_book} originate from interferences 
between partially scattered electronic waves. In contrast to classical transport, quantum transport is 
thus fundamentally influenced by scattering phases, and the transport properties of electronic nanodevices 
operating at low temperatures are determined by complex transmission amplitudes, instead of real 
transmission probabilities. However, while only the squared modulus of the transmission appears in  
the Landauer-B\"uttiker formula for the conductance,\cite{Landauer70,Buttiker86} transmission phases 
themselves cannot be directly measured.

\begin{figure}
\centerline{\includegraphics[width=0.6\linewidth]{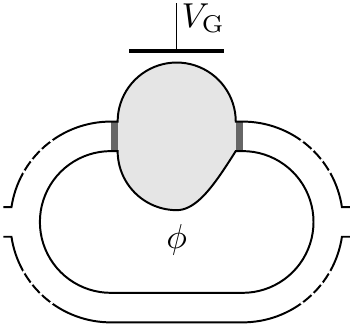}}
\caption{\label{Fig:ring_sketch}
Aharonov-Bohm interferometer, threaded by a flux $\phi$, with a quantum dot embedded
in its upper arm. The dot is capacitively coupled to a plunger gate potential $V_\mathrm{G}$,
which tunes its electrochemical potential, and thus changes its electronic occupancy. 
The unsymmetrical shape of the dot reflects the presumed lack of geometric symmetry
in experimental dots.
The dark segments at the entrances of the dot represent tunnel barriers.
The dashed lines on the arms of the interferometer stand for a number of possible additional leads.}
\end{figure}

In their pioneering phase-sensitive experiments, Yacoby \textit{et al.} \cite{Yacoby95} measured the
conductance oscillations of an AB interferometer with a quantum dot (QD) embedded in one of its arms. 
The QD operates in the Coulomb blockade (CB) regime and therefore only a single transverse channel 
participates in transport. The transmission of this channel through the QD is characterized by the complex 
amplitude $t=|t|e^{i \alpha}$.
Varying the voltage $V_\mathrm{G}$ on a nearby plunger gate capacitively coupled to the QD allows for the addition of 
electrons one by one, and the phase of the AB conductance oscillations can be monitored as a function of 
$V_\mathrm{G}$.
In the two-terminal setup of Ref.\ [\onlinecite{Yacoby95}], sketched in Fig. \ref{Fig:ring_sketch}, 
the conductance reads
\begin{equation}
\begin{aligned}
g_\mathrm{AB}(V_\mathrm{G},\phi)&=g_\mathrm{AB}^{(0)}(V_\mathrm{G})\\
&+\sum_p g_\mathrm{AB}^{(p)}(V_\mathrm{G}) \cos{\left(2\pi p \frac{\phi}{\phi_0} + \beta_p(V_\mathrm{G}) \right)},
\label{eq:AB}
\end{aligned}
\end{equation}
where $\phi$ is the flux through the AB ring and $\phi_0=hc/e$ is the flux quantum. The conductance
exhibits AB oscillations with $V_\mathrm{G}$-dependent characteristic phases $\beta_p$. An Onsager 
reciprocity relation dictates that $g$ is an even function of $\phi$ in a two-terminal setup.\cite{Buttiker86} 
Thus, $\beta_p(V_\mathrm{G})=0,\pi$ are the only two possible values.\cite{levy95} In the experiment of
Ref.\ [\onlinecite{Yacoby95}], $\beta_1$ was monitored, and abrupt jumps between those two  values were observed at 
values of $V_\mathrm{G}$ corresponding to CB resonances, i.e. where an electron is added on the QD. 
Assuming that $\beta_1$ is directly related to the transmission phase $\alpha$, such jumps could be
explained by Friedel's sum rule. More puzzling, however, were the additional, equally abrupt jumps of $\pi$ 
systematically observed in the CB  conductance valleys in between each and every two consecutive CB resonances. 

Two fundamental questions have been raised at that point. First, what is the connection between the conductance 
phases $\beta_p$ and the transmission phase $\alpha$? In other words, under which conditions is it possible to extract
the transmission phase $\alpha$ from the experimentally measurable phases $\beta_p$?
Second, what is the physical mechanism responsible for the in-phase behavior, 
i.e. the systematic $\pi$-jumps observed in between CB resonances?

It was understood early\cite{levy95} that the two-terminal setup had to be abandoned to probe the 
transmission phase $\alpha$. Opening the system to more terminals lifts the reciprocity constraints 
and allows for a one-to-one correspondence between $\alpha$ and $\beta_1$.
This was experimentally achieved by Schuster \textit{et al.} \cite{Schuster97} who opened the arms of the 
interferometer to additional grounded terminals -- this is sketched by the dashed lines in 
Fig.\ \ref{Fig:ring_sketch}.
Working with such a ``leaky" interferometer suppresses processes with multiple windings around the ring, 
so that only $\beta_1$ can be extracted. An appropriate tuning of the opening of the ring arms in this 
multiterminal setup allows for the identification of $\beta_1$ with $\alpha$,\cite{Aharony02,EntinWohlman02}
and Ref.\ [\onlinecite{Schuster97}] obtained the expected Breit-Wigner behavior of the phase, with a smooth
increase of $\pi$ every time a CB resonance is crossed.
However, the second puzzle of Ref.\ [\onlinecite{Yacoby95}] persisted, as the systematic phase lapses of 
$\pi$ in-between any pair of consecutive CB resonances also appeared in the multiterminal setup. 

Refs.\ [\onlinecite{Yacoby95}] and [\onlinecite{Schuster97}] both work with hundreds of electrons on the QD.
The experiments were performed at very low temperatures ($T \approx 80$ mK) where the temperature was estimated
to be the smallest energy scale in the system.
In a more recent experiment at even lower temperatures ($T \approx 30$ mK), Avinum-Kalish \textit{et al.} \cite{Avinum05} 
investigated small QD's with zero to few tens of electrons. Their estimations of the energy scales involved in the experiments are 
the following: the temperature $K_bT \approx 0.003$ meV, the level spacing $\Delta \approx 0.5$ meV, the level width 
$\Gamma \approx 0.03-0.3$ meV, and the charging energy $U=1-3$ meV. Their key observation is that, as the number of electrons on the QD is 
reduced from 20 down to 0, $\beta_1$ undergoes a crossover from the \textit{universal phase behavior} regime, 
with regularly alternating $\pi$ jumps at and in-between CB resonances, to a \textit{mesoscopic regime} where 
phase lapses in-between CB resonances occur in a random fashion, so that the in-phase behavior of the 
transmission at CB resonances gets lost.
These experiments provide an important hint towards the resolution of the puzzle: candidate theories applicable in the
experimental regime ($T < \Gamma < \Delta < U$ ) have to 
be able to explain the universal to mesoscopic crossover as the QD is depopulated.

This seminal series of works motivated further experiments. The role of the magnetic field
was explored by Sigrist \textit{et al.} in a AB ring with one QD embedded in each of its arms.
\cite{Sigrist04} The phase of a QD in the Kondo regime was measured by Ji \textit{et al.}
\cite{Ji00,Ji02} Highly controlled experiments coupled with detailed theoretical modeling
for an AB device without QD also found phase lapses for
some parameter values due to scattering and reflections in the arms of the ring.\cite{Kreisbeck10}

The puzzles posed by the experimental data attracted a sustained theoretical interest.
Refs.\ [\onlinecite{Wu98,Kang99,Aharony02,EntinWohlman02}] established that, under not too restrictive 
constraints, $\alpha$ can be extracted from $\beta_1$ in multi-terminal geometries.
Assuming therefore that $\beta_1=\alpha$, Refs.\ [\onlinecite{levy00,taniguchi99,Lee99,kim03}]
enounced the simple rules that in non- (or weakly) interacting systems, $\pi$ phase jumps occur under the two 
following circumstances: (i) the electrochemical potential crosses an eigenmode of the scatterer (so that 
one electron is added to the QD); (ii) the transmission $t$ vanishes.
For non- (or weakly) interacting systems, the universal regime of transmission phase thus implies that
there is one transmission zero in-between any two consecutive CB resonances. Numerical simulations on a 
non-interacting disordered diffusive lattice model\cite{levy00} however showed that transmission zeros 
occur in-between consecutive resonances with probability close to $1/2$. From this result it is often 
concluded that noninteracting theories are unable to explain the experimental data.

A mechanism based on level occupation switching initially considered in 
Refs.\ [\onlinecite{Hackenbroich97,Baltin99a}] has been repeatedly used within non-interacting \cite{Oreg07} and 
interacting models.\cite{Silvestrov00,Silvestrov07,Goldstein09} 
A Fano-like scenario is assumed to stem from a given (broadened) QD level which is much more strongly coupled 
to the leads than all nearby (narrow) levels. Then, while the position of CB resonance peaks is determined 
by the energy of the narrow levels, the conductance is dominated by the transmission through the broadened 
level. As $V_\mathrm{G}$ is varied, the narrow levels are successively populated right after the CB resonance, 
at which point the broadened level is abruptly depopulated. It was initially argued that level occupation 
switching arises from specific spatial structures of the QD,\cite{Hackenbroich97,Baltin99a} but later works 
went further and suggested that it generically follows from electronic 
correlations.\cite{Silvestrov00,Silvestrov07,Goldstein09,Kim06,Golosov06,bertoni07} 

\begin{figure}
\centerline{\includegraphics[width=\figwidth]{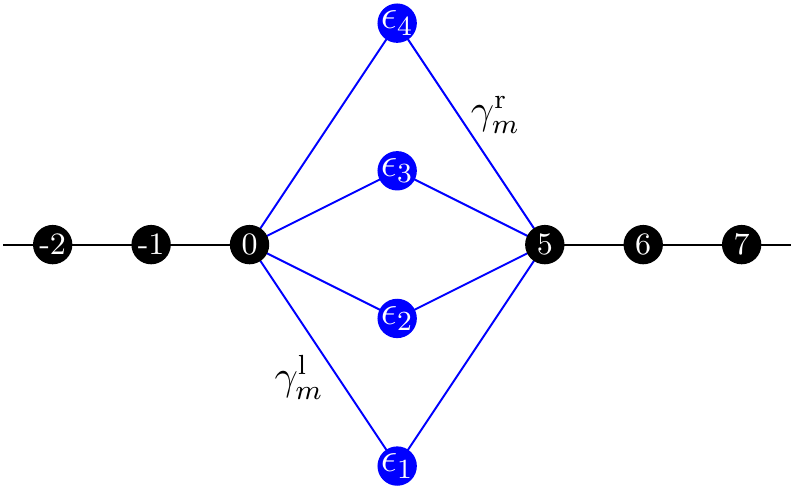}}
\caption{\label{fig:RLMsketch} (Color online) Sketch of the topology of the Resonant Level Model.}
\end{figure}
More recently, lattice models for interacting fermions were numerically investigated,
and an interpretation was proposed in which the electronic correlations induce the mode switching mechanism.
Karrasch \textit{et al.} investigated few-level, strongly interacting systems where the in-phase behavior is 
obtained when the single-particle level spacing on the QD becomes smaller than the level broadening
due to the coupling to the leads.\cite{Karrasch07,Karrasch2}
Varying these parameters (and the interaction strength) allowed to drive the transition between the universal 
and mesoscopic regime within a given dot of fixed (small) size and number of electrons, unlike the experimental 
case where the transition is obtained by changing the electron number and thus the filling of the dot.
The mode-switching mechanism becomes relevant when
the electronic population is large enough to justify a mean-field treatment of interactions.
This conclusion is somehow at odds with the numerical results of Ref.~[\onlinecite{levy00}],
given that a mean-field approximation essentially delivers a single-particle theory. 

Bergfield \textit{et al.} considered strongly correlated models of molecules, where the universal behavior 
of the phase cannot be reached, unless spatial symmetries are imposed on the molecule itself and the 
molecule-lead couplings.\cite{Bergfield10} This latter result indicates that the universal regime requires 
irregular single-particle spectra, very different from regular molecular orbital spectra.
Gurvitz proposed that the phase behavior in the transmission through a quantum dot 
results from the formation of a Wigner molecule.\cite{Gurvitz08}

In another line of work, a very simple solution to the puzzle posed by 
Refs.~[\onlinecite{Yacoby95,Schuster97,Avinum05}] was recently proposed.\cite{Molina12} The approach is based on the 
Constant Interaction Model (CIM), which treats QD in the CB regime as noninteracting, up to a constant charging 
energy term. The sole assumption that wave-functions have quantum chaotic spatial correlations\cite{berry77} is 
able to reproduce the two main experimental observations of (i) long,
universal sequences of in-phase resonances, and (ii) a crossover to a mesoscopic regime for not too large 
number of electrons on the QD. The probability of deviating from the universal behavior can be obtained as a 
function of the electron filling. Given the success of the CIM in describing different features of CB 
physics,\cite{Jalabert92,AlhassidRMP2000} it was expectable that the statistical behavior of the transmission 
phase were also within its reach.    

\begin{figure}
\centerline{\includegraphics[width=\figwidth]{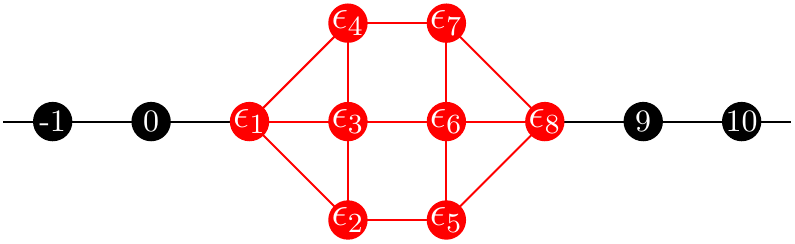}}
\caption{\label{Fig:LMsketch}
(Color online) Lattice model for a QD with $M=8$ sites (red) connected to two
external one-dimensional leads (black).}
\end{figure}
To bring the experimental-theoretical controversy to closure, it is nevertheless important to develop and investigate
more realistic models incorporating electron-electron interactions beyond the CIM (in particular including strong 
correlations), and check if they also can reproduce the experimental observations, for instance via the level 
occupation switching mechanism. This is one of the main goals of this work. In particular, we are interested in 
knowing if the occurrence of transmission zeros obtained with particular models of interacting electrons can 
account for a genuine universal regime with in-phase behavior over long sequences of resonances.
For that purpose we use two different, generic models of QD: the \textit{resonant level model} (RM), 
describing a finite number of single-particle states with repulsively interacting electrons, 
individually connected to external Fermi liquid leads (see Fig.~\ref{fig:RLMsketch}),
and \textit{lattice models} (LM) with nearest-neighbor electronic repulsion, connected to two Fermi liquid leads via 
two sites (see Fig.~\ref{Fig:LMsketch}).
Our investigations are restricted to the experimentally relevant regime of zero temperature linear response 
and we do not address the extreme cases of models with only one or two resonant levels nor situations in which 
Kondo physics is relevant.\cite{Ji00,Ji02,Karrasch2}

We use the DMRG-based embedding method to extract the transmission properties of the 
system\cite{molina03,molina04,MeSch03} and its recent extension to calculate the transmission 
phase.\cite{Molina12a} We review such an approach in Appendix~\ref{Sec:embedding} and illustrate its 
power for a one-dimensional interacting quantum wire, including a numerical verification of the 
Friedel sum rule in Appendix~\ref{Sec:appb}. 

The manuscript is organized as follows.
In Sec.\ \ref{Sec:1plattice} we introduce the one-particle version of the lattice and resonant models 
that we use throughout the paper. Following the standard literature, we relate the transmission zeros 
to a matrix resolvent and to the distribution of the coupling widths, which allows to identify the 
different situations in which a transmission zero can appear. 
Although such an analysis is not directly applicable when correlations are present, 
it can be extended to the many-particle case whenever an effective single-particle theory can be constructed.
Furthermore, we will present numerical evidences that the behavior observed in one-particle models often 
also applies to strongly correlated cases. The many-particle RM is defined and studied 
in Sec.\ \ref{Sec:mprlm}. Results in the limit of large resonance widths are shown, where new zeros appear 
but no in-phase behavior of neighboring resonances is achieved. The next Sec.\ \ref{Sec:mplm} is devoted 
to many-particle lattice models, where examples are shown for different sizes and ratios between the one-particle 
level spacing and the level width. 
No significant trend towards universal behavior emerges upon inclusion of correlation effects.
Finally, some conclusions and perspectives are discussed in Sec.\ \ref{Sec:conclusions}.

\section{Transmission zeros in one-particle models}
\label{Sec:1plattice}

Lattice models of a one-particle QD connected to one-dimensional leads were first used by Levy-Yeyati and 
B\"uttiker\cite{levy00} for the numerical calculation of the transmission phase. The one-particle 
resonant level model, where the eigenstates of the QD are linked to one-dimensional leads through 
hopping amplitudes, was solved 
in Ref.\ [\onlinecite{Aharony02}] and the necessary conditions for the appearance of transmission 
zeros were determined. These two models have been extremely useful for discussing the behavior of 
the transmission phase in the non-interacting case, as well as when electron-electron interactions are 
treated within the CIM or at the mean-field level. 
In addition, both models can be further generalized to fully account for interactions by adding 
a Coulomb term coupling the non-interacting basis states. This is the route that we follow in Sections 
\ref{Sec:mprlm} and \ref{Sec:mplm}, and thus we start by presenting the main concepts concerning 
one-particle models that will be later generalized to describe the interacting case. 
Moreover, our discussion of one-particle models will help to systematize and classify the 
various approaches previously proposed to analyze the existence of transmission zeros.  

An arbitrarily-shaped, non-interacting QD with $M$ sites, connected to one-dimensional leads through its 
1\textsuperscript{st} and $M$\textsuperscript{th} sites, as sketched in Fig.\ \ref{Fig:LMsketch}, 
is generically described by the Hamiltonian
\begin{equation}
H=H_\mathrm{D}+H_\mathrm{G}+H_\mathrm{C}+H_\mathrm{L}.
\label{Eq:generalH}
\end{equation}
The left and right lead Hamiltonians are given by
\begin{equation}
H_\mathrm{L} = - \sum_{i=-\infty}^0 \left( c^{\dagger}_{i+1} c^{\phantom{\dagger}}_{i} + h.c. \right)
- \sum_{i=M+1}^{\infty} \left( c^{\dagger}_{i+1} c^{\phantom{\dagger}}_{i} + h.c. \right) \, .
\label{Eq:Hleads}
\end{equation}
$c_i^{\phantom{\dagger}}$ ($c_i^{\dagger}$) indicate the standard operators for 
annihilation (creation) of a spinless fermion on site $i$ and \textit{h.c.} stands for the 
Hermitean conjugate. 
We have chosen throughout this work the hopping amplitude in the leads to be the unit of energy. 

Within the lattice model, the QD Hamiltonian reads
\begin{equation}
H_\mathrm{D}^\mathrm{LM}
=-t_\mathrm{D} \sum_{\left< i j \right>} \left( c^{\dagger}_j c^{\phantom{\dagger}}_i + h.c. \right)
+\sum_{i=1}^M \epsilon_i n_i,
\label{Eq:HDlattice}
\end{equation}
where $\left< i j \right>$ denotes a pair of nearest neighbor lattice sites and 
$n_i= c^{\dagger}_i c^{\phantom{\dagger}}_i$. The hopping amplitude within the dot $t_\mathrm{D}$ can be different 
from the one in the leads, and we use such a freedom in Sec.\ \ref{Sec:mplm}. In the absence of magnetic field 
the hopping amplitudes can be chosen to be real without loss of generality. Disorder is modeled by random on-site 
energies $\epsilon_i$, taken from a uniform distribution of width $W$. 

The capacitive coupling to a nearby gate is proportional to the total number of electrons in the QD,
\begin{equation}
\label{Eq:Hg}
H_\mathrm{G}=-V_\mathrm{G}\sum_{i=1}^M n_i \, ,
\end{equation} 
where capacitances are included into the definition of $V_\mathrm{G}$.

Since only the sites $i=1$ and $i=M$ are connected to the leads, the coupling Hamiltonian between the leads 
and the QD is
\begin{equation}
H_\mathrm{C}^\mathrm{LM}= -t_\mathrm{C} 
\left( c^{\dagger}_0 c^{\phantom{\dagger}}_1+ c^{\dagger}_1 c^{\phantom{\dagger}}_0 \right)
-t_\mathrm{C} \left(c^{\dagger}_{M+1} c^{\phantom{\dagger}}_M+ c^{\dagger}_M c^{\phantom{\dagger}}_{M+1}\right).
\label{Eq:HClattice}
\end{equation}
The left and right hopping amplitudes connecting the leads to the QD could in principle be different from that of 
the leads in order to achieve the regime of weak coupling. However, we will restrict ourselves to  $t_\mathrm{C}=1$ 
throughout this work. The choice of one-dimensional leads is justified since in the tunneling regime characterizing 
CB physics, only a single transverse mode per lead is relevant.\cite{AlhassidRMP2000}

Having defined our one-particle lattice model, we next follow Ref.\ [\onlinecite{levy00}] and discuss the 
conditions under which a transmission zero occurs between two resonances. Our starting point is the retarded 
Green's function of the QD, which can be written as
\begin{equation}
G(\epsilon)=(\epsilon-H_\mathrm{D}-\Sigma(\epsilon))^{-1},
\end{equation}
with the self-energy $\Sigma(\epsilon)$ arising from the coupling to the leads.
The transmission amplitude 
is related to the Green's function via the Fisher-Lee
relation \cite{FisherLee}
\begin{equation}
t(\epsilon)=i \hbar (v_\mathrm{l} v_\mathrm{r} )^{1/2} G_{1M}(\epsilon),
\label{Eq:LeeFisher}
\end{equation}
where $v_\mathrm{l(r)}$ is the velocity in the left (right) lead for electrons with energy $\epsilon$.
When leads are connected to a single QD site, 
the condition for having $t=0$ can be written as \cite{levy00}
\begin{equation}
{\cal C}_{1M} \left\{\epsilon-H_{D}+\Sigma(\epsilon)\right\}=0
\label{Eq:cofactor}
\end{equation}
where ${\cal C}_{1M}\{A\}$ stands for the cofactor of the $(1,M)$ matrix element of $A$. Since $\Sigma_{11}$ 
and $\Sigma_{MM}$ are the only non-zero elements of the matrix $\Sigma$, the condition expressed in
Eq.~(\ref{Eq:cofactor}) is entirely determined by the properties of the isolated QD and is thus independent of 
the coupling strength to the leads. This important observation allows us to locate the transmission zeros from 
those of the matrix element 
\begin{equation}\label{eq:resolvent}
{\cal F}_{1M}(\epsilon)= \sum_m \frac{\psi_m(1) \psi_m(M)}{\epsilon-\epsilon_m},
\end{equation}
of the resolvent for the isolated QD, where $\epsilon_m$ and $\psi_m(1)$($\psi_m(M)$) are the $m^{\rm th}$
QD's eigenenergy and eigenfunction evaluated at site $1(M)$, respectively. 

The structure of ${\cal F}_{1M}(\epsilon)$ is characteristic of physical situations where resonances are coupled 
to a continuum, and allows to determine the existence of zeros according to the residues of the poles. As we 
will see, the resonant model, that we present below, leads to equivalent conditions.

The one-particle resonant level model is obtained after a basis transformation of the QD's degrees of freedom 
from the site basis to the QD's eigenbasis. The Hamiltonian in Eq.~\eqref{Eq:generalH} retains the same structure, 
with however new QD and coupling terms,
\begin{eqnarray}
H_\mathrm{D}^\mathrm{RM} &=& \sum_{m=1}^{M} \epsilon_{m} n_m \, ,
\label{EQ:HDRM} \\
H_\mathrm{C}^\mathrm{RM} &=&\sum_{m=1}^M \left( \gamma_m^\mathrm{l} 
d^{\dagger}_m c^{\phantom{\dagger}}_0  + \gamma_m^\mathrm{r}  d^{\dagger}_m c_{M+1}  + h.c. \right) \, .
\label{Eq:HcoupRM}
\end{eqnarray}
The new fermionic operators $d_m = \sum_n \psi_m(n) c_n$ are obtained from the old ones via a unitary 
transformation with the eigenfunctions $\psi_m$ of $H_\mathrm{D}^\mathrm{LM}$, and $n_m = d^\dagger_m d_m$. 
In this way the levels $\epsilon_m$ are understood as eigenvalues of an isolated QD. Alternatively they can 
be interpreted as the on-site energies of a tight-binding model with the topology of Fig.\ \ref{fig:RLMsketch} 
and hopping amplitudes given by the partial-width amplitudes $\gamma_m^\mathrm{l,r}$. The total widths are 
$\Gamma_m=|\gamma_m^\mathrm{l}|^2+|\gamma_m^\mathrm{r}|^2$.

An exact solution for the transmission amplitude in this model was presented in Ref.\ [\onlinecite{Aharony02}], 
which obtained
\begin{equation}
t(\epsilon)=\frac{2 i \, f_{\mathrm{lr}}  \sin{k}}
{(f_{\mathrm{ll}}+e^{-ik})(f_{\mathrm{rr}}+e^{-ik})-|f_{\mathrm{lr}}|^2},
\label{Eq:tRLM}
\end{equation}
where $k$ is the wave-vector in the lead and
\begin{equation}
f_\mathrm{xy}(\epsilon)=\sum_{m} \frac{\gamma^{\mathrm{x}}_{m} \gamma^{*\mathrm{y}}_{m}}{\epsilon-\epsilon_{m}},
\label{Eq:SLR}
\end{equation}
for $\mathrm{x},\mathrm{y}=\mathrm{l},\mathrm{r}$. With one-dimensional leads the partial amplitudes are 
proportional to the values of the resonant wave-functions at the extreme points. Therefore, 
$f_\mathrm{lr}(\epsilon)$ is simply proportional to the function $\mathcal{F}_{1M}(\epsilon)$ of Eq.\ 
(\ref{eq:resolvent}) and to the $R$-matrix element of the corresponding scattering 
problem.\cite{AlhassidRMP2000,Jalabert92}

When there are only small variations in the values of the wave-functions on the sites connecting to the leads, 
the behavior of ${\cal F}_{1M}(\epsilon)$ away from $\epsilon_m$ is dictated by the two surrounding singularities. 
On the other hand, large fluctuations of $\psi_m(1) \psi_m(M)$ among different $m$ might lead to values of 
${\cal F}_{1M}(\epsilon)$ determined by far-away resonances. These two possible situations will be respectively 
referred to as \textit{Restricted Off-Resonance} (ROR) behavior and \textit{Unrestricted Off-Resonance} (UOR) 
behavior. This distinction plays a key role within the analysis of transmission 
phases.\cite{Hackenbroich97,Baltin99a,kim03,Silvestrov00,Silvestrov07} Since large wave-function fluctuations in 
generic systems are rare, we will find that the ROR is the most commonly encountered scenario. Interestingly, the 
above classification is not only relevant for the one-particle models, but also for many-particle models 
(Secs.\ \ref{Sec:mprlm} and \ref{Sec:mplm}).

The existence of a transmission zero between the $m$\textsuperscript{th} and the $(m+1)$\textsuperscript{st}
resonances depends on the sign of \cite{levy00,Silva02}
\begin{equation}\label{signrule}
D_m=\psi_m(1) \psi_m(M)  \psi_{m+1}(1) \psi_{m+1}(M) \, .
\end{equation}
In the ROR case, when $D_m > 0$ (equal parity of the resonances) there is one zero between the resonances 
while for $D_m<0$ (opposite parity) there is no zero. This {\it sign rule} has been at the basis of several studies of 
the transmission phase. In the UOR case we have that for $D_m > 0$ there is an odd number of transmission zeros 
between the $m$\textsuperscript{th} and $(m+1)$\textsuperscript{st} resonances, yielding an accumulated phase of 
$\pi$ between the two resonances. For $D_m < 0$ there is no transmission zero or there is an even number of zeros,
resulting in zero total phase shift between the two resonances. In the interferometric experiments on QDs 
operating in the CB regime it is extremely difficult to follow the transmission phase across the conductance 
valleys, and only the total phase between resonances is relevant. Therefore, the sign rule is also useful in 
the UOR case. At this point it is important to remark that the occupation switching mechanism is based on a 
large fluctuation of the partial width leading to the UOR behavior. As we have seen, the appearance of 
transmission zeros when $D_m < 0$ is indeed possible in the UOR case, but they are bound to come in pairs, 
without altering the in-phase relationship of adjacent resonances.

\begin{figure}
\hspace{9mm}\includegraphics[width=\figwidthb]{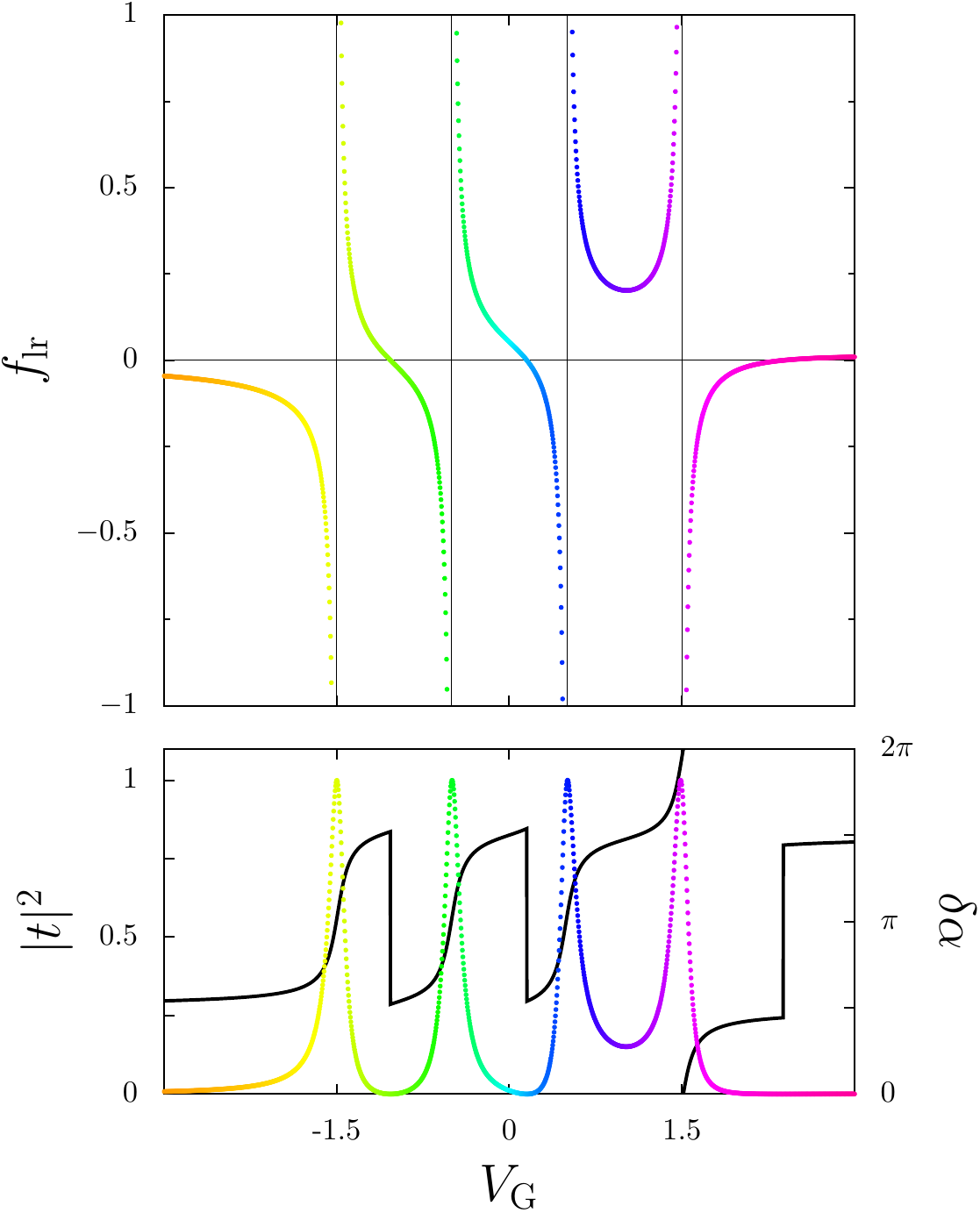}\\[2mm]
\includegraphics[width=\figwidth]{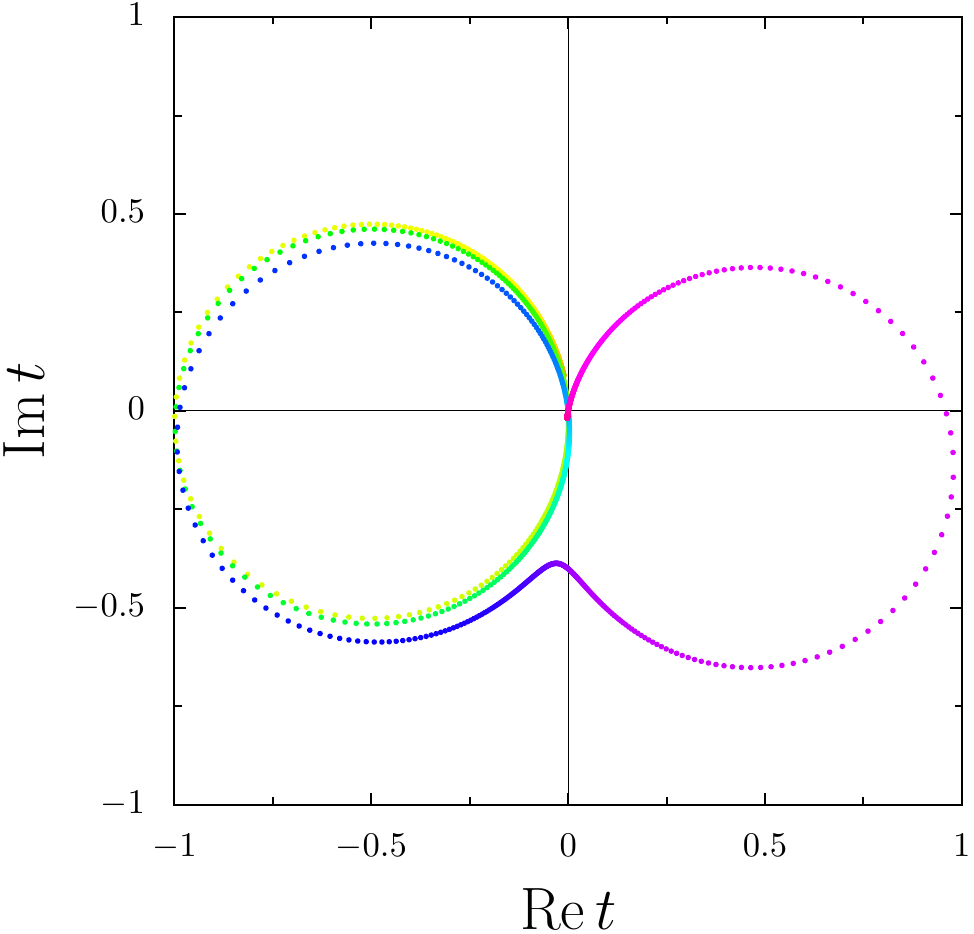}
\caption{\label{Fig:Example1} 
(Color online) Numerically obtained data for  a four-resonance one-particle resonant level model with
$\epsilon_m=m-2.5$, $\gamma^\mathrm{l}_m=\gamma^\mathrm{r}_m=0.2$ for $m=1,2,3$ and
$\gamma^\mathrm{l}_4=-\gamma^\mathrm{r}_4=0.2$, exhibiting ROR behavior in all intervals. 
Top panel:
$f_\mathrm{lr}$ (defined in Eq.\ \ref{Eq:SLR}).
Middle panel:
Transmission coefficient $|t|^2$ (colored data points) and phase $\alpha$ (solid black line) as a function 
of $V_\mathrm{G}$. 
Bottom panel: 
trajectories of the transmission amplitude in the complex plane when $V_\mathrm{G}$ is varied as on the 
other panels. The color rainbow scale gives the correspondence between data in different panels.}
\end{figure}

\begin{figure}
\hspace{9mm}\includegraphics[width=\figwidthb]{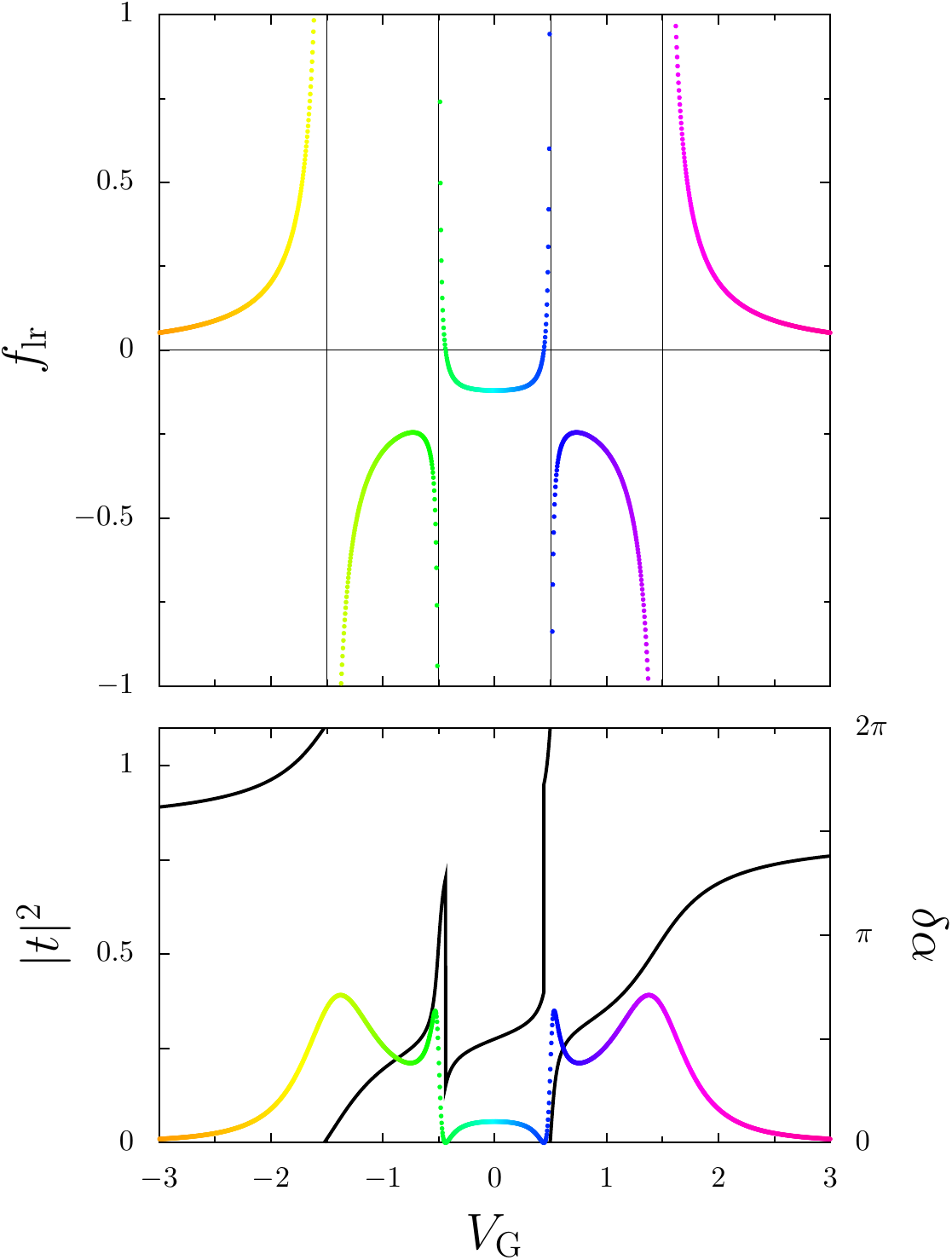}\\[2mm]
\includegraphics[width=\figwidth]{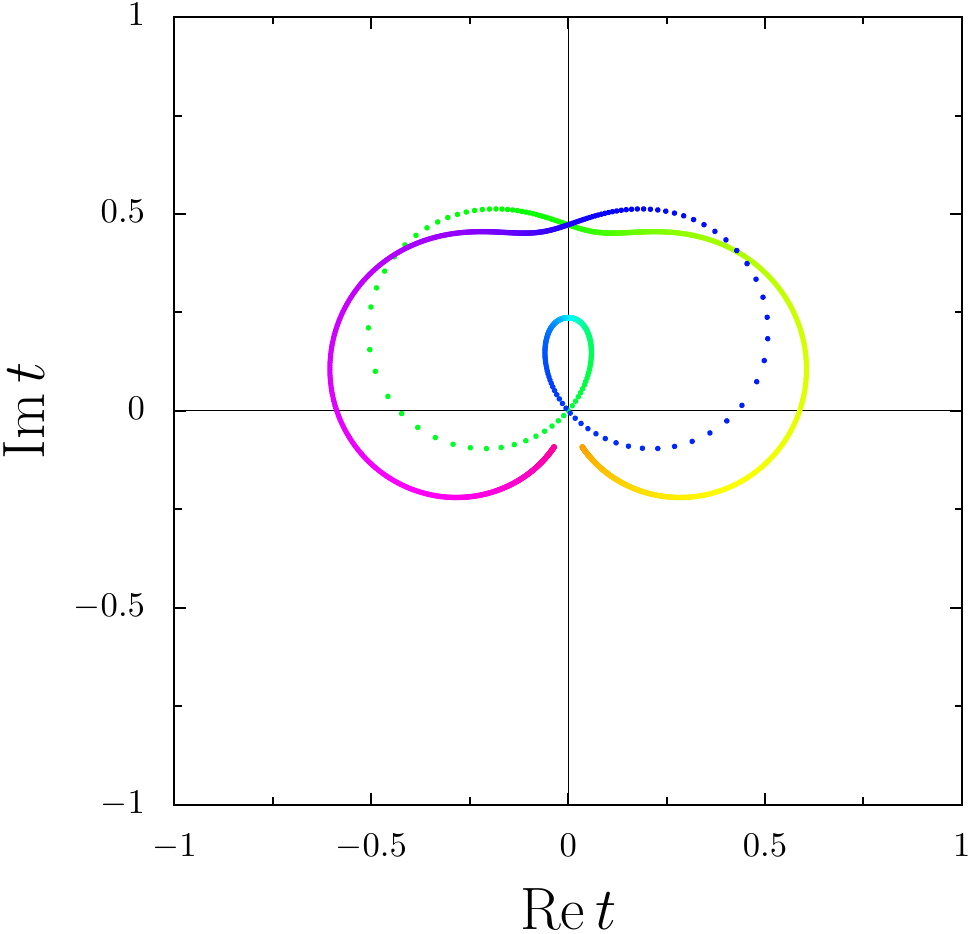}
\caption{\label{Fig:Example2} 
(Color online) Numerically obtained data for  a four-resonance one-particle resonant level model with
$\epsilon_m=m-2.5$, 
$\gamma_1^{\mathrm{l}}=\gamma_4^{\mathrm{l}}=-0.6$, 
$\gamma_2^{\mathrm{l}}=\gamma_3^{\mathrm{l}}=\gamma_4^{\mathrm{r}}=-0.2$, $\gamma_2^{\mathrm{r}}=-0.05$,
$\gamma_3^{\mathrm{r}}=0.05$, and $\gamma_1^{\mathrm{r}}=0.2$, exhibiting UOR behavior between the 
2\textsuperscript{nd} and 3\textsuperscript{rd} resonance. The same conventions as in Fig.\ \ref{Fig:Example1} 
are adopted.}
\end{figure}
We next discuss the ROR-UOR competition and the sign rule in the one-particle RM. The experience gained in 
this simple one-particle situation will prove very useful for understanding the many-particle results in the 
next sections. Compared to the LM, the RM has the advantage that partial-width coupling amplitudes 
$\gamma_m^\mathrm{l,r}$ can be tuned at will. This flexibility has been a key ingredient in several theoretical 
works.\cite{Baltin99a,Silvestrov00,Silvestrov07,Golosov06,Karrasch07,Karrasch2}

We first show in Fig.~\ref{Fig:Example1} results for a system with four equidistant resonances and coupling 
parameters $\gamma^\mathrm{l}_m=\gamma^\mathrm{r}_m=0.2$, for all $m=1,2,3,4$, with the exception of 
$\gamma^\mathrm{r}_4=-0.2$. The values of the couplings all have the same magnitude, thus leading to ROR behavior. 
The upper panel of Fig.\ \ref{Fig:Example1} shows the resulting function $f_\mathrm{lr}$ when varying the gate 
voltage $V_\mathrm{G}$ (which is equivalent to varying the energy $\varepsilon$ of the incoming particles 
from the lead) in arbitrary units. The middle panel of Fig.\ \ref{Fig:Example1} shows the transmission 
coefficient $|t|^2$ and the phase $\alpha$ as a function of $V_\mathrm{G}$. We can see that the presence and 
the absence of zeros between resonances is clearly correlated with the sign of $D_m$.
Transmission resonances and zeros correspond, respectively, to singularities and zeros of $f_\mathrm{lr}$.
The trajectory of the transmission amplitude in the complex plane as a function of $V_\mathrm{G}$ is shown in 
the bottom panel. The trajectories crossing the origin of the complex $t$-plane occupy only one
half-plane for the first three resonant peaks, and are characteristic of adjacent resonant peaks with 
in-phase behavior.\cite{taniguchi99} Even in this simple example there is an anomalous extra transmission 
zero at $V_\mathrm{G} \approx 2.3$ far away from the area of resonance peaks. This is a finite-size effect, 
arising from zeros of $f_\mathrm{xy}$ that lie outside the interval where the resonances are concentrated. 
Such a behavior is irrelevant for CB experiments with many resonances, and we only note that, according to 
our nomenclature, it is an UOR behavior, as the value of the transmission far from resonance is dominated by 
the contribution from several resonant tails. Such an effect is commonly encountered in finite size numerical 
simulations and it is discussed for instance in Ref.[\onlinecite{Karrasch07}].

Typical UOR behavior is shown in Fig.\ \ref{Fig:Example2}, where we reduced the values of the two central 
resonance widths to make the behavior of $f_\textrm{lr}$ between the second and third resonances
dominated by the external resonances. A loop is formed by the trajectory in the complex plane and two 
new zeros appear. These zeros add phase lapses smaller than $\pi$ in the region between two resonances.
The peak appearing around $V_\mathrm{G}=0$ is related to a small loop close to the origin of the $t$-plane 
and an accumulated phase smaller than $\pi$. Therefore, it does not represent a new resonance. Instead, it is the 
result of a broad resonance being cut by the extra Fano-like anti-resonances. We will see that similar 
behaviors are also found in the many-particle case where the universal behavior is not reached.

A critical UOR-ROR case arises when the two zeros of $f_\mathrm{lr}$ collapse in a double zero (with a 
horizontal tangent) yielding a kink at the origin of the complex $t$-plane with a phase shift smaller than $\pi$
(not shown). This special zero is found for specific settings of the parameters and if the couplings are
perturbed infinitesimally away from that singular setting, we find either no zero or two zeros in the 
transmission. 

The inclusion of interactions at the CIM level opens a gap of the size of the charging energy in the 
one-particle spectrum between the highest occupied and the lowest unoccupied dot levels without affecting 
the wave-functions. The UOR behavior will then be favored since many resonances contribute in the electron and hole 
sectors.\cite{Baltin99b} However, as explained above, the sign rule dictating the phase behavior only concerns 
the two nearest resonances. It is therefore important to consider more refined models including arbitrary couplings 
and electronic correlations. We undertake this task in the forthcoming sections and discuss the results in 
connection with previously proposed theories and the concepts introduced in this Section.

\section{Many-particle resonant level model}
\label{Sec:mprlm}

\begin{figure}
\hspace{9mm}\includegraphics[width=\figwidthb]{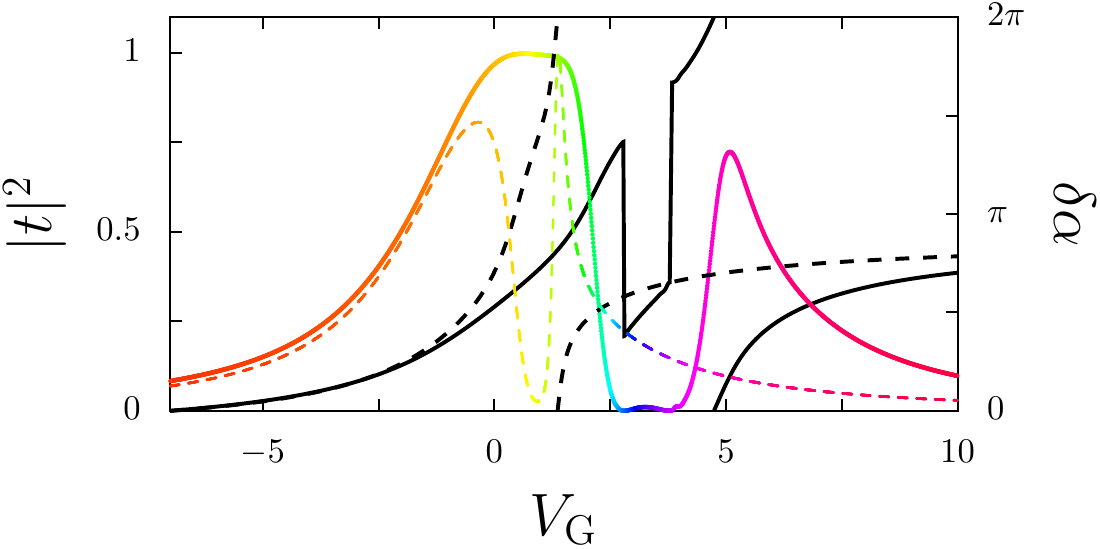}\\[2mm]
\includegraphics[width=\figwidth]{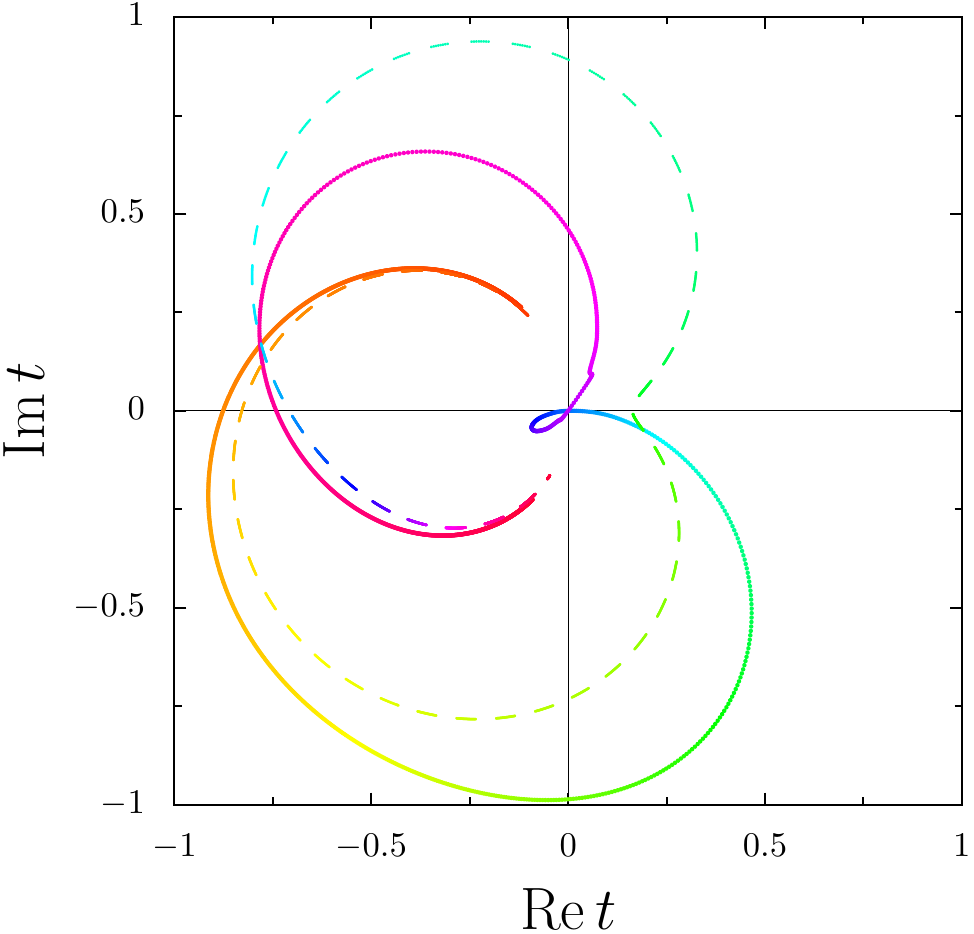}
\caption{\label{fig:fano3} (Color online) Upper panel: transmission coefficient (colored) 
and phase (black) for non-interacting ($U=0$, dashed) and interacting ($U=2$, solid)
resonance level models with three levels and
parameters $\epsilon_1=-0.5$, $\epsilon_2=0.5$, $\epsilon_3=1.5$, $\gamma_1^\mathrm{l}=1.0$,
$\gamma_1^\mathrm{r}=1.0$, $\gamma_2^\mathrm{l}=-0.5$, $\gamma_2^\mathrm{r}=0.5$,
$\gamma_3^\mathrm{l}=0.4$, $\gamma_3^\mathrm{r}=0.7$
Lower panel: trajectory of the transmission in the complex plane following the convention of the last three figures.}
\end{figure}
The one-particle resonant level model is particularly useful to investigate the possibility of UOR behavior, 
because partial widths can be tuned at will. Since interactions at the CIM level do not favor the universal 
behavior of the phases, it is natural to ask whether correlation effects beyond mean field could be responsible 
for the experimentally observed universal behavior. 
This question has been analyzed by Karrasch \textit{et al.}, \cite{Karrasch07} via numerical investigations of
many-particle resonant level models with up to $M=4$ levels and arbitrary level-lead couplings.
They identified several regimes, determined by three energy scales: the mean level-width $\Gamma$,  the mean 
one-particle level spacing $\Delta$ in the QD, and the strength of the electron-electron interactions $U$.
In the case $\Gamma \lesssim \Delta$, interactions were not found to affect the behavior of transmission phases. 
In contrast, when $ U\simeq \Gamma > \Delta$, sufficiently strong correlations appeared to favor the appearance 
of additional transmission zeros not predicted by the sign rule.
This result was interpreted as the signature of Fano-like antiresonances between a renormalized wide resonance 
with several narrow single-particle levels, thus providing some justification for the level occupation switching 
mechanism. While the phase lapses reported in Ref.~[\onlinecite{Karrasch07}] seem consistent with the universal 
regime, it is not so obvious that the zero-temperature gate-voltage dependence of the transmission reproduces the 
experimentally observed standard CB resonances obtained from the amplitudes of the AB oscillations in the universal
regime. This is due to an incomplete filling of the resonances in the theoretical model 
resulting in an associated phase accumulation smaller than $\pi$, as we demonstrate below.

To better understand correlation effects on the transmission phase we reproduced and extended the results of 
Ref.\ [\onlinecite{Karrasch07}], working with up to $M=6$ levels in the QD. For $M=2$ and 4 our numerical 
calculations using the embedding technique (with Density Matrix Renormalization Group calculations) 
are in very good quantitative agreement with those obtained in 
Ref.\ [\onlinecite{Karrasch07}] with the Numerical Renormalization Group and the Functional 
Renormalization Group algorithms. The QD Hamiltonian of the many-particle resonant model is that of 
Eq.\ \eqref{EQ:HDRM} plus an interaction term
\begin{equation}
\label{Eq:HDrlint}
H_\mathrm{DU}^\mathrm{RM} =\frac{U}{2}
\sum_{m \ne m'} \left(n_m-\frac{1}{2}\right ) \left(n_{m'}-\frac{1}{2} \right) \, .
\end{equation}

In Fig.\ \ref{fig:fano3} we show that interactions can generate transmission zeros in dots with 
$\Gamma > \Delta$ already for $M=3$ states. The fluctuations of the level-widths are small enough that the 
non-interacting case (dashed lines) exhibits pure ROR behavior. According to the sign rule, and since there are 
alternating parities between consecutive resonances, there is no transmission zero. The broad peak (around 
$V_\mathrm{G}=-1$) of the non-interacting case (top panel, dashed lines) is the result of the overlap of two 
nearby resonances. In the interacting case with $U=2$ (top panel, solid lines), three peaks 
(around $V_\mathrm{G}=0$, 3, and 5) can be seen. Two zeros are clearly identified from the phase lapses of 
$\alpha$ and from the trajectories of $t$ in the complex plane (bottom panel). The small peak between the two 
zeros corresponds to a small loop in the complex $t$-plane, indicating the emergence of UOR behavior induced 
by interactions. The increase of $\alpha$ through each of the peaks is smaller than $\pi$, which would be the 
expected value if one electron were added to the QD.

Our example is consistent with the mechanism put forward by Karrasch \textit{et al.} [\onlinecite{Karrasch07}],
where the transmission amplitude in the vicinity of a new zero is not determined by the two nearest resonances but 
by an anomalously broadened level. However, the transmission phase change from one of the new zeros to the next 
one does not correspond to the full addition of an electron on the QD. As a direct consequence, the 
$V_\mathrm{G}$-dependence of the conductance diverges from the typical CB peak structure. We observed 
similar behavior in all the realizations where extra zeros appeared with interactions. The analysis of these simple 
cases illustrates the usefulness of the discussion presented in the previous section. Even if the function 
$f_\mathrm{lr}(\varepsilon)$ of Eq.\ (\ref{Eq:SLR}) does not have meaning in a many-body situation, the concepts of 
ROR and UOR behavior can be addressed by varying $V_\mathrm{G}$ and studying the resulting trajectory in the 
complex $t$-plane. 

\begin{figure}
\hspace{9mm}\includegraphics[width=\figwidthb]{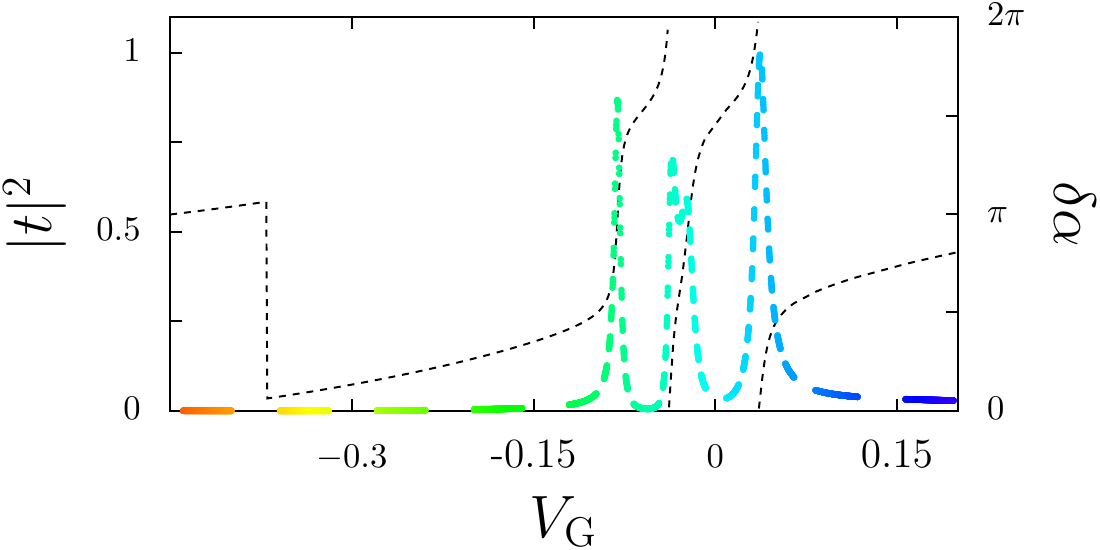}\\
\hspace{9mm}\includegraphics[width=\figwidthb]{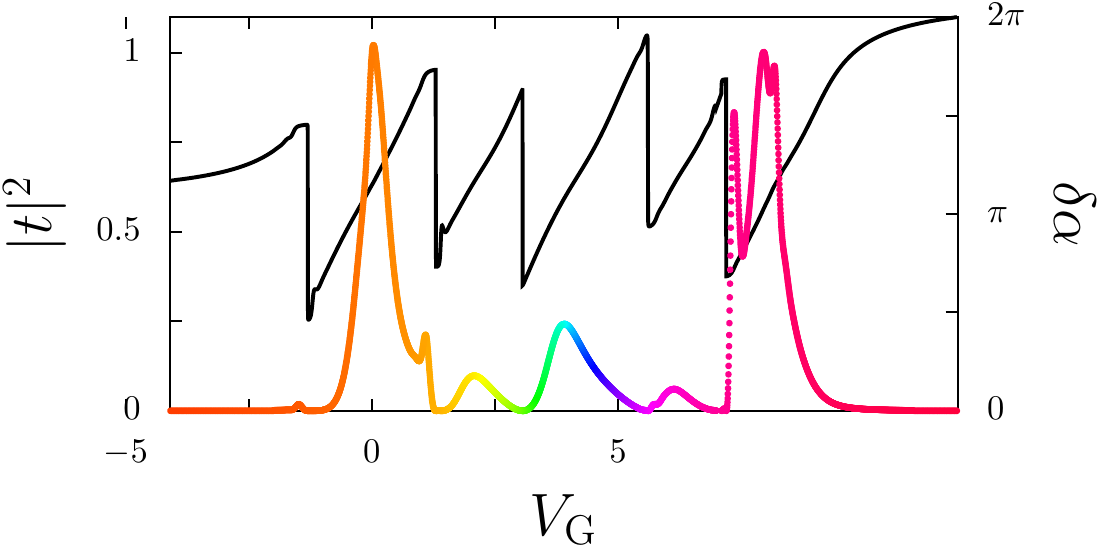}\\[2mm]
\includegraphics[width=\figwidth]{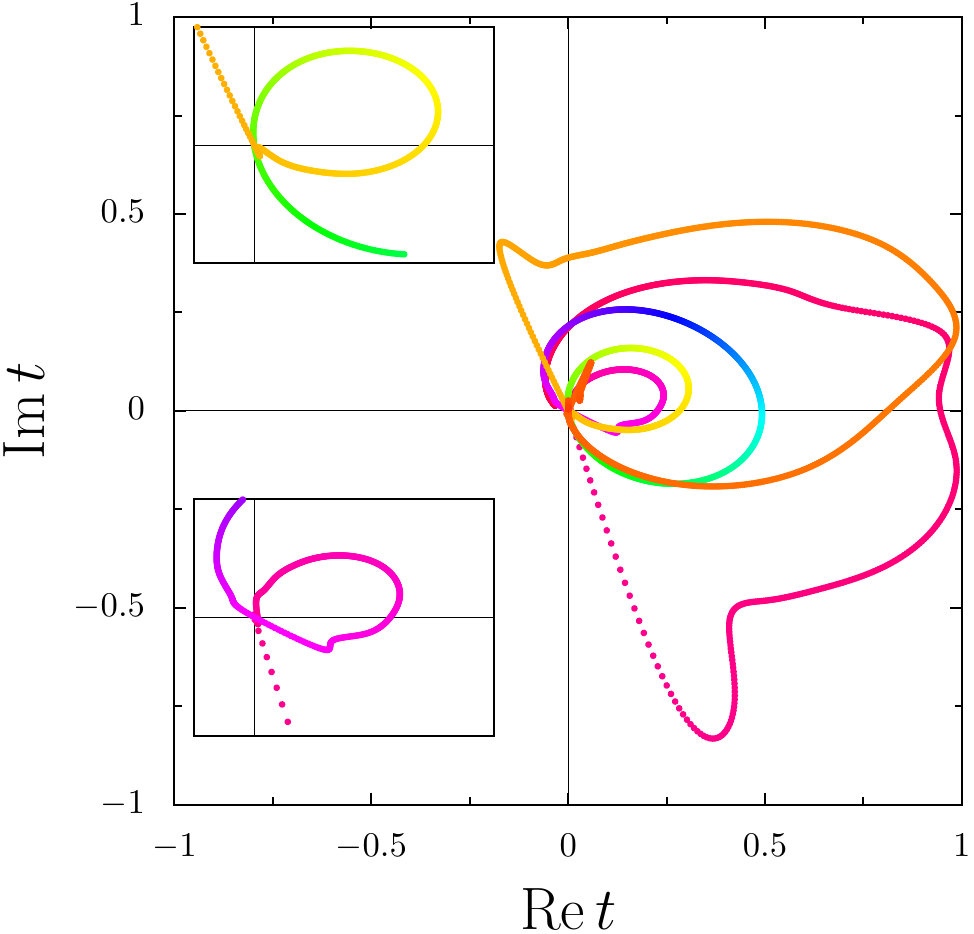}
\caption{\label{fig:s6gaml} (Color online) Transmission coefficient (color) and transmission 
phase (black) of a RM with six levels for $U=0$ (upper panel, dashed) and $U=2$ (center and lower panels, 
solid lines and data points). The model parameters are  
$\epsilon_1=-0.113$, $\epsilon_2=-0.071$, $\epsilon_3=-0.059$, $\epsilon_4=0.004$, 
$\epsilon_5=0.016$, $\epsilon_6=0.082$,
$\gamma_1^\mathrm{l}=0.3883$,
$\gamma_1^\mathrm{r}=0.1775$, $\gamma_2^\mathrm{l}=-0.2106$, $\gamma_2^\mathrm{r}=0.3292$,
$\gamma_3^\mathrm{l}=0.2448$, $\gamma_3^\mathrm{r}=0.1214$, $\gamma_4^\mathrm{l}=0.2581$, 
$\gamma_4^\mathrm{r}=-0.1936$, $\gamma_5^\mathrm{l}=0.3155$, $\gamma_5^\mathrm{r}=0.2332$,
$\gamma_6^\mathrm{l}=0.4253$, $\gamma_6^\mathrm{r}=-0.1977$. 
The ratio between the average energy spacing and the average level width is $\Delta/\Gamma=0.26$. 
Bottom panel: 
trajectories of the transmission amplitudes in the complex plane as a function of $V_\mathrm{G}$. The insets show 
$V_\mathrm{G}$-intervals where transmission zeros occur together with overlapping resonances.} 
\end{figure}
The advantage of our DMRG-based embedding method is that it allows to increase the QD size to larger values than 
previously studied, still keeping a good precision. We next extend the above analysis to larger systems, to 
investigate the possible crossover between the mesoscopic and the universal regime of the transmission phase.  
In all examples to be discussed, we randomly chose the widths $\gamma_m^\mathrm{l(r)}$, while tuning the 
values of the resonance energies $\epsilon_m$ in order to achieve different behavior in particular examples.

\begin{figure}
\hspace{9mm}\includegraphics[width=\figwidthb]{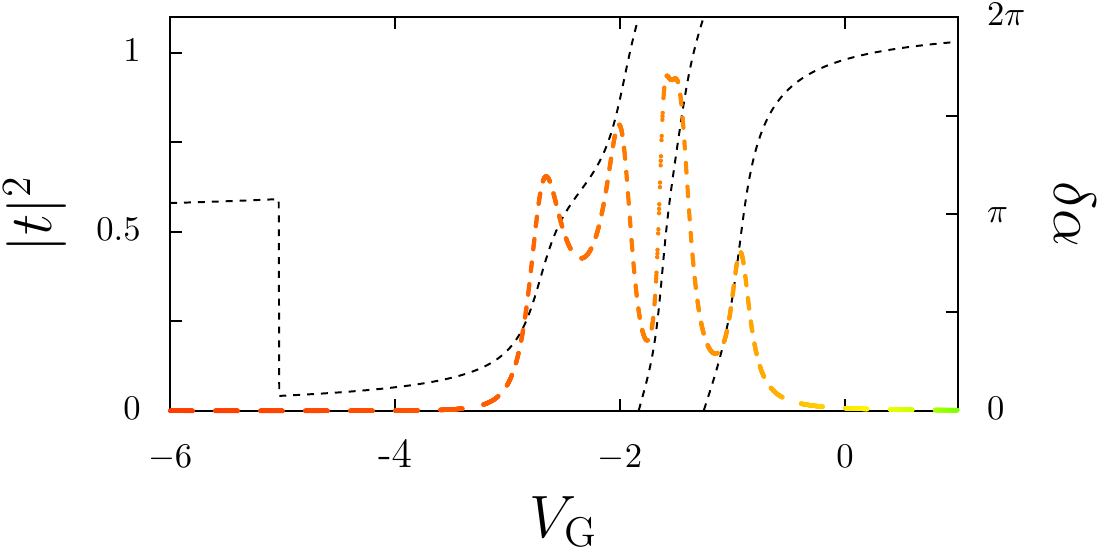}\\
\hspace{9mm}\includegraphics[width=\figwidthb]{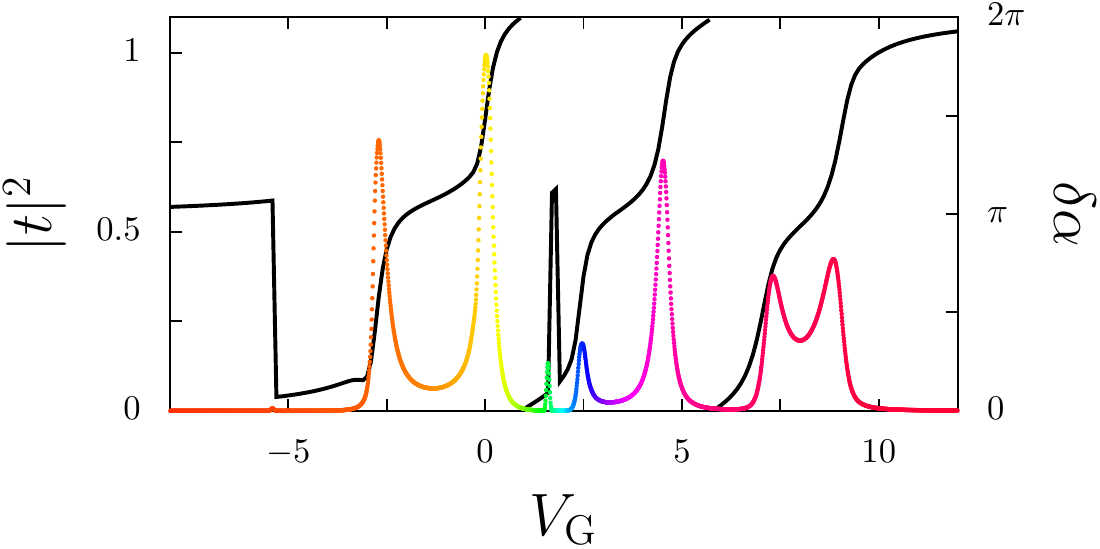}\\[2mm]
\includegraphics[width=\figwidth]{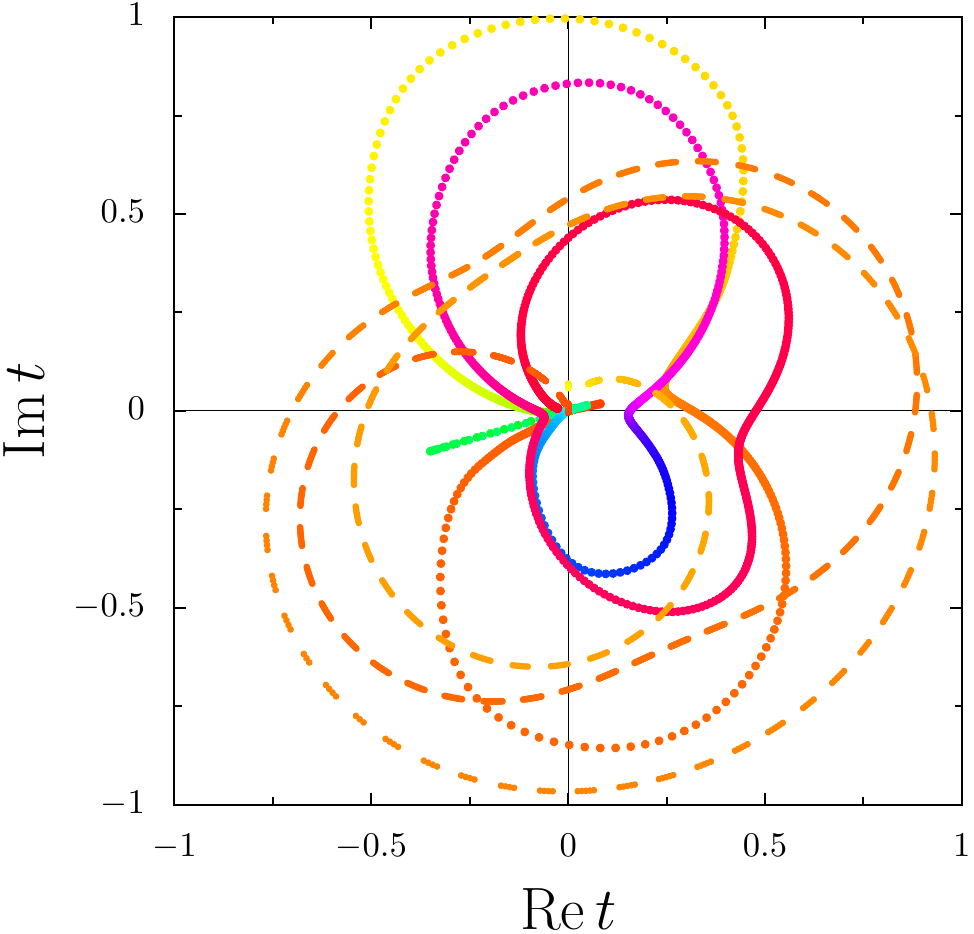}
\caption{\label{fig:fano6U2} (Color online) Transmission coefficient (color) and transmission phase (black)  
of a RM with six levels for $U=0$ (upper panel, dashed) and $U=2$ (central panel, solid).
The coupling amplitudes are the same as in Fig.\ \ref{fig:s6gaml}, while the one-particle energies are  
$\epsilon_1=-1.7497$, $\epsilon_2=-1.0161$, $\epsilon_3=-0.6483$, $\epsilon_4=-0.4684$, 
$\epsilon_5=0.0428$, and $\epsilon_6=0.16212$, giving $\Delta/\Gamma=2.16$. 
Bottom panel: Trajectory of the transmission amplitude in the complex plane as a function 
of $V_\mathrm{G}$.}
\end{figure}
We show in Fig.\ \ref{fig:s6gaml} data for a QD with $M=6$ resonant levels, where the average energy spacing is 
much smaller than the average level width $\Delta \ll \Gamma$. The fluctuations of the coupling amplitudes are not 
strong enough to induce UOR behavior in the non-interacting case. The sign rule in the ROR case dictates the 
absence of transmission zeros between the resonances (upper panel), and there is one zero outside the region of 
the resonances. For $U=2$ (central panel) there is one transmission zero outside the region of the resonances and 
four zeros in the region of the resonances. In the same way as the previous example of a smaller QD, the phase 
evolution does not exhibit in-phase behavior from one resonance peak to the next, due to the non-integer filling 
of the dot at each resonance. The trajectory of the transmission in the complex $t$-plane shows loops typical of 
UOR behavior. For instance, the two resonances close to $V_\mathrm{G}=0$ are associated with a phase increase 
larger than $\pi$, and are ``cut" by a zero at $V_\mathrm{G}\simeq 1$. The insets in the lower panel show loops 
between two zeros that are characteristic of conductance peaks that do not represent a resonance with the 
corresponding integer filling of the dot. The phase evolution in the loop is then smaller than $\pi$, and the 
angle of the crossing of the trajectories at the origin directly gives the missing dot filling. 

\begin{figure}
\hspace{9mm}\includegraphics[width=\figwidthb]{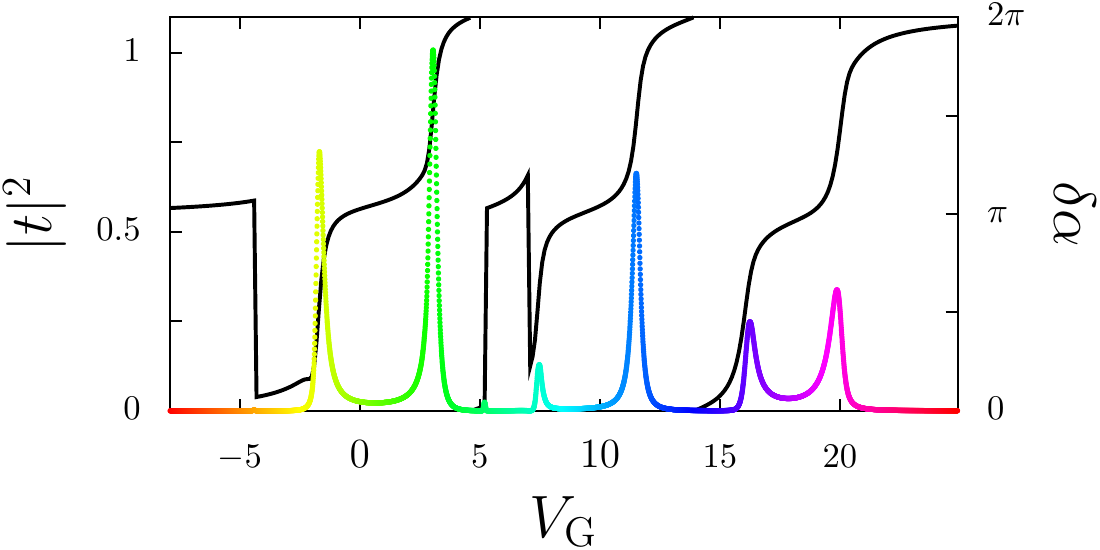}\\[2mm]
\includegraphics[width=\figwidth]{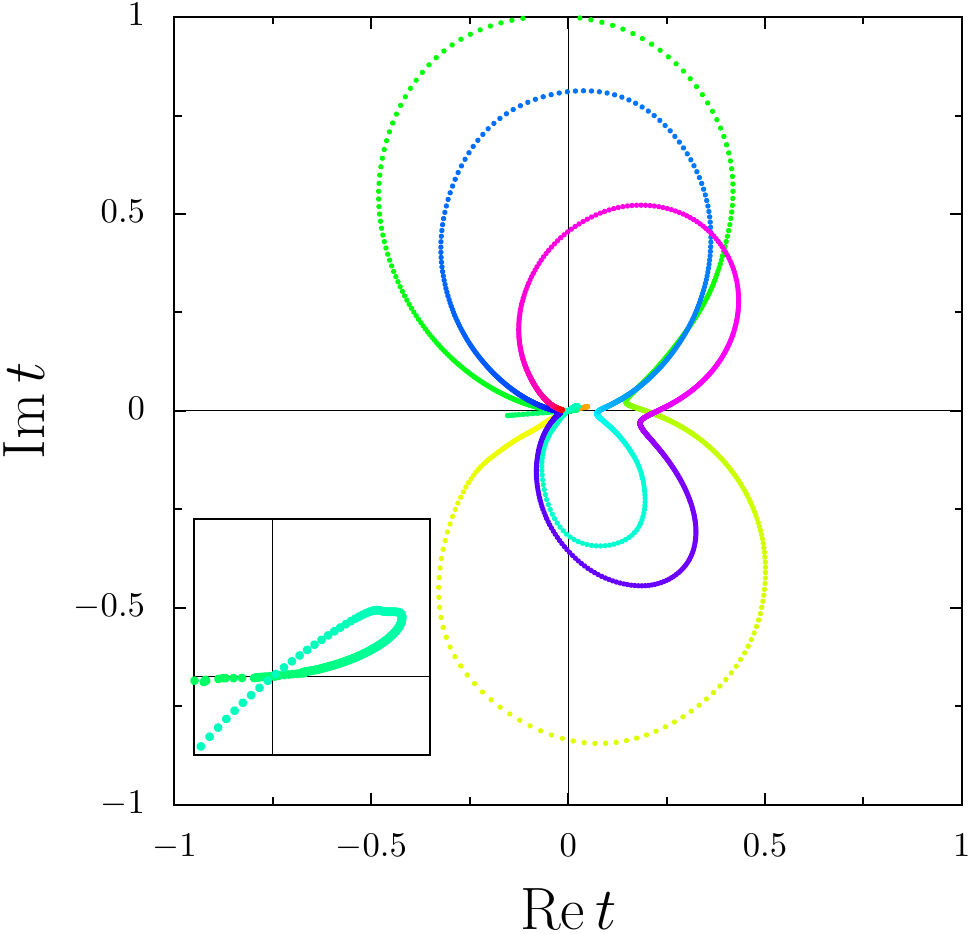}
\caption{\label{fig:fano6U4} (Color online) Results for a resonance level model with six levels and the same 
parameters as in Fig.\ \ref{fig:fano6U2}, at the interaction strength $U=4$.
Top panel: Transmission coefficient (colored line/data points) and phase (black line).
Bottom panel: Trajectory of the transmission in the complex plane as a function of $V_\mathrm{G}$.
The inset shows the $V_G$ interval where the transmission zeros occur together with overlapping resonances.}
\end{figure}
We next explore the emergence of UOR behavior by taking different ratios of $\Gamma/\Delta$ within the 
various resonances of a given sample, considering the influence of the interaction strength $U$. We present 
results for a QD with $M=6$ resonant levels, where the energy spacing between the upper two levels is much 
smaller than all other spacings for different interaction strengths: $U=0$ and 2 (Fig.\ \ref{fig:fano6U2}) 
and $U=4$ (Fig.\ \ref{fig:fano6U4}). For $U=0$ (dashed lines) we are in the regime of overlapping resonances. 
This is clearly indicated by the transmission probability in the top panel of Fig. \ref{fig:fano6U2}. There is 
an anomalous transmission zero outside the resonance area, arising from the same mechanism as discussed in 
Sec.\ \ref{Sec:1plattice}. The sample shows ROR behavior at $U=0$, and in agreement with the sign rule, there is 
no transmission zero. When the interactions are turned on, the charging energy leads to the separation of 
resonances, which evolve into CB peaks. Resonances no longer overlap, except the last two, whose energy 
separation in the non-interacting case is particularly small. The change in the behavior of the trajectory of 
the transmission amplitude in the complex plane induced by the effect of the interactions is remarkable. 
Two new zeros appear rather close to each other between the second and the third resonance thus exhibiting 
UOR behavior with a vanishing accumulated phase shift (see insets of the lower panels). They are located in 
the region where the phase evolution is driven by the beginning of the filling of the third level. 
No new zeros appear between the most overlapping last two resonances. As the interaction is increased further 
the qualitative behavior and the number of zeros is not changed. We show in Fig.\ \ref{fig:fano6U4} the results 
for stronger interaction strength, $U=4$, where the main difference is that CB peaks become narrower and more 
distant. The two zeros between the second and the third CB peak also become more separated by the effect of 
the increasing interaction. We checked for $U=6$ and 8 (not shown) that, except for this trivial effect,
the trajectories of $t$ in the complex plane almost do not change with $U$.

\begin{figure}
\hspace{9mm}\includegraphics[width=\figwidthb]{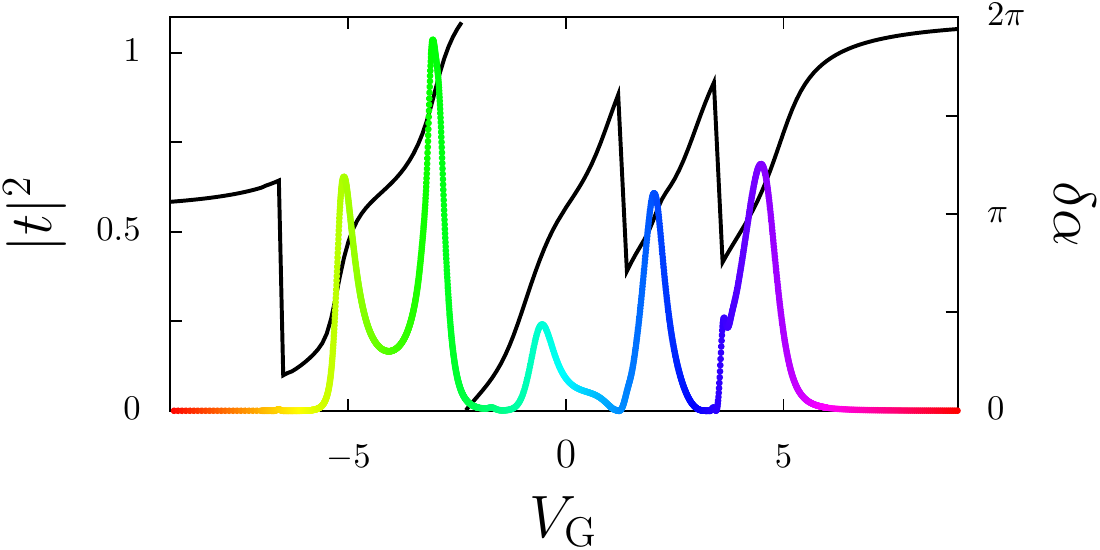}\\[2mm]
\includegraphics[width=\figwidth]{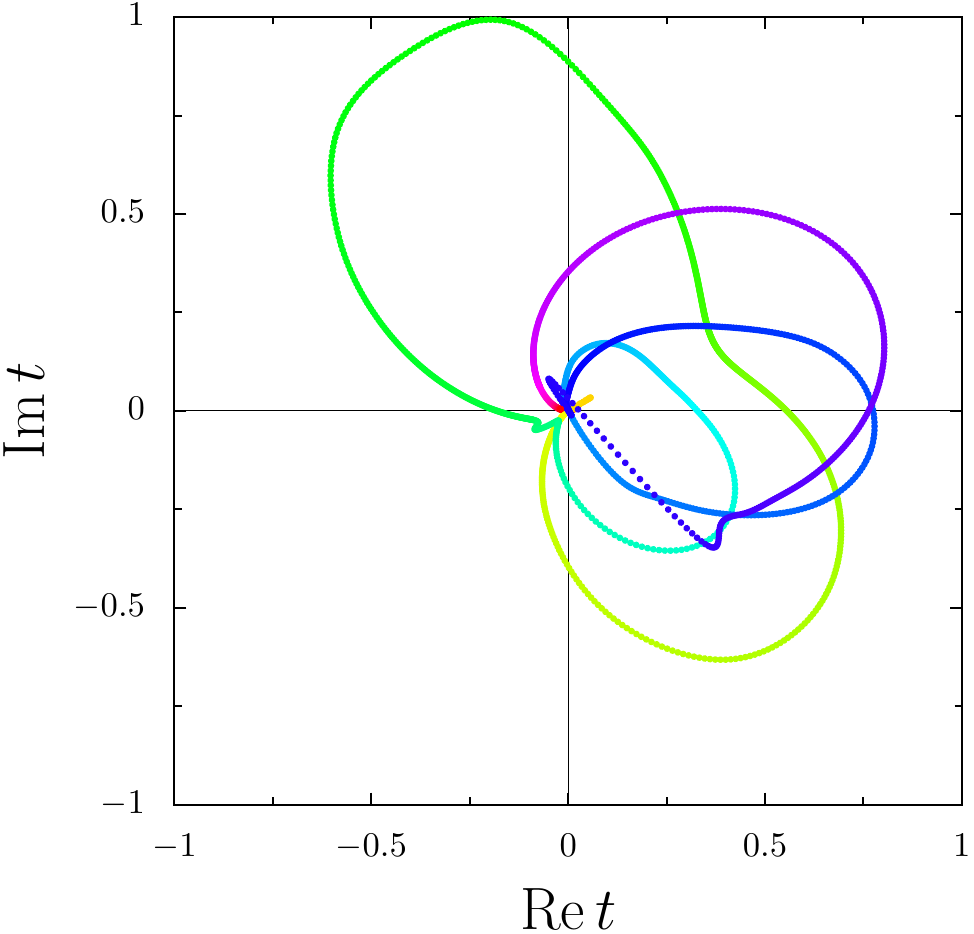}
\caption{\label{fig:fano6b} (Color online)
Top panel: Transmission coefficient (colored line) and phase (black line) 
for a resonance level model with six levels and $U=2$. The coupling amplitudes are the same as in 
Fig.\ \ref{fig:s6gaml}, while the one-particle energies are
$\epsilon_1=-0.44976$, $\epsilon_2=-0.2161$, $\epsilon_3=-0.04683$, $\epsilon_4=0.0042$, 
$\epsilon_5=0.01621$, and $\epsilon_6=0.018$, giving $\Delta/\Gamma=0.635$. 
Bottom panel: Trajectory of the transmission in the complex plane as a function of $V_\mathrm{G}$.}
\end{figure}

In a last example of the resonant model, we kept the same couplings as in the previous examples, while significantly 
reducing the level spacings to enter the regime $\Delta < \Gamma$. 
We see in Fig.\ \ref{fig:fano6b} that the number of zeros does not change with respect to the interacting cases of 
Figs.\ \ref{fig:fano6U2} and \ref{fig:fano6U4}, but we can obtain very small loops in the $t$-plane, resulting 
from the extreme UOR behavior. 

The previously discussed examples show that the extrapolation towards larger systems by 
increasing $\Gamma/\Delta$ is delicate, and large variations of this ratio are needed 
in order to generate transmission zeros between each pair of consecutive resonances. 

From our numerical results in the many-particle RM we conclude that (i) level occupation switching induced 
by interactions appears only in the extreme $\Delta \ll \Gamma$ case where the zeros can cut through 
resonances, and (ii) it is directly related to UOR behavior where the transmission between resonances is 
not simply given by the independent contributions from the nearest two resonances. 
There is a wide intermediate regime $\Delta \approx \Gamma$ where UOR behavior appears only 
for part of the resonances. The interaction must satisfy $U \gg \Delta$ and $U \gtrsim \Gamma$ 
for the effect to appear,\cite{Karrasch07} but once this is the case, increasing $U$ further does 
not induce any qualitative changes in the number of transmission zeros and 
the associated phase lapses. Finally, the occurrence of this mechanism should translate into a 
\textit{partial occupation of the dot} when the special zeros induced by correlations appear. 
In this case, the form of the conductance peaks seems to depart significantly depart from the well resolved
CB peaks observed in the universal regime.

\section{Many-particle lattice models}
\label{Sec:mplm}

While the RM is very useful to discuss the various possible scenarios that may lead to the appearance of 
transmission zeros and the associated phase lapses, the single-level energies and half-width amplitudes 
are independent parameters of the model. We therefore return to lattice models in this section, where the 
single-level parameters are determined by the geometry of the lattice system.

The dot Hamiltonian for a lattice model of spinless fermions with nearest-neighbor interaction is given by that 
of Eq.\ \eqref{Eq:HDlattice} plus the interaction term
\begin{equation}
\label{Eq:HDlmint}
H_\mathrm{DU}^\mathrm{LM} = U \sum_{\left<ij\right>} \left( n_i-\frac{1}{2} \right) 
\left( n_j-\frac{1}{2} \right).
\end{equation}
Such a nearest-neighbor repulsion can lead to strong correlations depending on the value of the interaction 
strength $U$. Writing the interaction term in the basis of the one-particle dot eigenstates, it would take the 
form of \eqref{Eq:HDrlint}, but with level-dependent interaction strengths and widths calculated from the dot
eigenfunctions. 
The limitation to nearest-neighbor interaction considerably simplifies the numerical work, but it is not of 
fundamental nature.

When $U \ne 0$, it is not possible to make {\em a priori} general statements about the behavior of the zeros. 
We therefore performed numerical calculations using the embedding method presented in detail in the 
App.\ \ref{Sec:embedding}. We considered systems ranging from the simplest topology of a diamond with $M=4$ sites
(Fig.\ 1 of Ref.\ [\onlinecite{Molina12a}]) to dot sizes up to $M=8$ sites (see Fig.\ \ref{Fig:LMsketch}). 

When the single-particle level-spacing is of the order of the level-width ($\Delta\gtrsim\Gamma$), the 
transmission zeros with $U\neq0$ typically exhibit the same qualitative behavior as in the non-interacting 
case $U=0$ [\onlinecite{Karrasch07,Molina12a}]. This situation typically happens in small systems when 
$t_\mathrm{D}=1$. For instance, in the case with $M=4$ and on-site energies chosen such that one has a 
transmission zero at $U=0$, the effect of the interactions is to separate the peaks, reducing their widths 
without changing the qualitative behavior of the scattering phase \cite{Molina12a}. 

\begin{figure}
\centerline{\includegraphics[width=\figwidth]{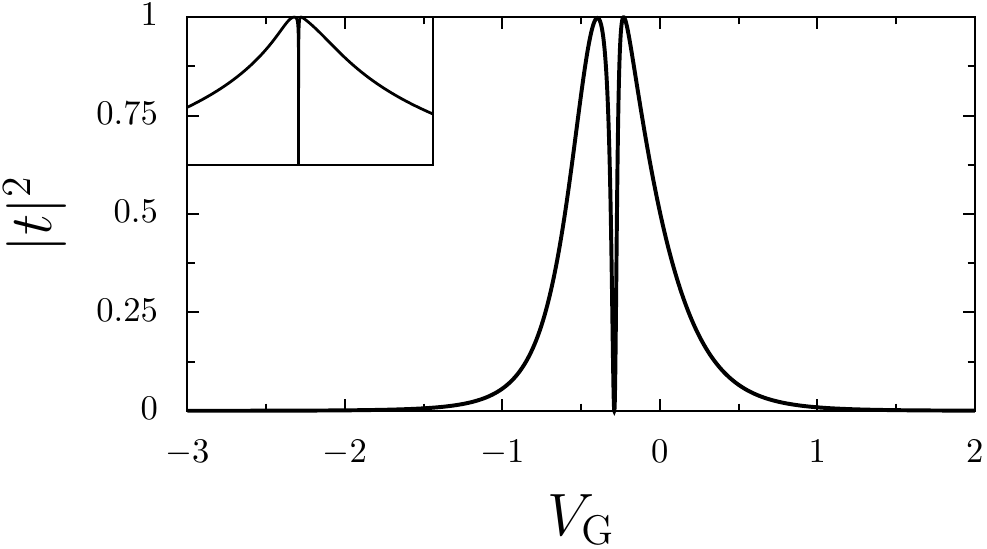}}
\centerline{\includegraphics[width=\figwidth]{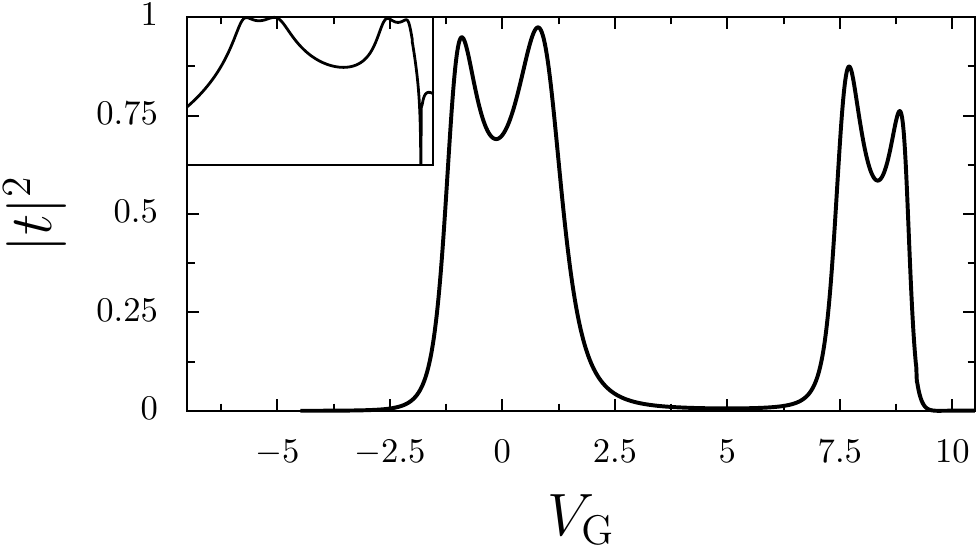}}
\caption{\label{fig:4site2} Transmission coefficient \textit{vs.} gate voltage of a lattice model for a quantum 
dot with four sites and $t_\mathrm{D}=0.25$ for $U=0$ ($4$) in top (bottom) panel. The chosen disorder has led 
to on-site energies $\epsilon_1=0.220$, $\epsilon_2=0.360$, $\epsilon_3=0.219$, $\epsilon_4=-0.037$. The 
logarithmic scale used in the insets allows to locate the transmission zeros.}
\end{figure}
In simple models of small dots without disorder, interactions do not modify the number of transmission zeros from 
the non-interacting case and they do not induce a change from ROR to UOR behavior. By varying the on-site energies 
and taking $t_\mathrm{D}<1$, we occasionally obtain the displacement of a transmission zero from the region between 
the resonances to the zone outside the resonances. Such a behavior is shown in Fig.\ \ref{fig:4site2}, for the case 
$M=4$, $t_\mathrm{D}=0.25$, $U=4$ and a realization of the on-site energies $\epsilon_i$ randomly chosen in the 
interval $[-W/2,W/2]$ with $W=1$. In this realization, there is a transmission zero between two groups of 
resonances in the case $U=4$; $t_\mathrm{D}=1$ and also in the non-interacting case $U=0$ when $t_\mathrm{D}=0.25$. 
However, this zero disappears in the shown case of $t_\mathrm{D}=0.25$ and $U=4$, while another zero appears 
outside the resonance area. The disappearance can be related to a change in the wave function of the dot that 
evolves into a charge density wave in the interacting case.

According to Ref.\ [\onlinecite{Karrasch07}], the evolution from the mesoscopic to the universal regime can be 
achieved if $\Delta\lesssim\Gamma\lesssim U$. To check this idea, we reached the previous condition within the 
many-particle LM by going to relatively large systems ($M=8$) and taking the hopping within the dot smaller than 
in the leads.

Systems of different sizes ($M=4$ to $8$) level spacings ($t_\mathrm{D}=1$ and $t_\mathrm{D}<1$) and interaction strengths
($U=0$ to $8$), with and without disorder, have been explored.  The numerical investigations are extremely time consuming, 
which is why we base our analysis on $|t|$ instead of calculating the complex $t$.\cite{numericaldifficulty} 
Below we present some illustrative examples supporting our general conclusion:
interactions do not typically induce the transition from the mesoscopic to the universal regime.  

\begin{figure}
\centerline{\includegraphics[width=\figwidth]{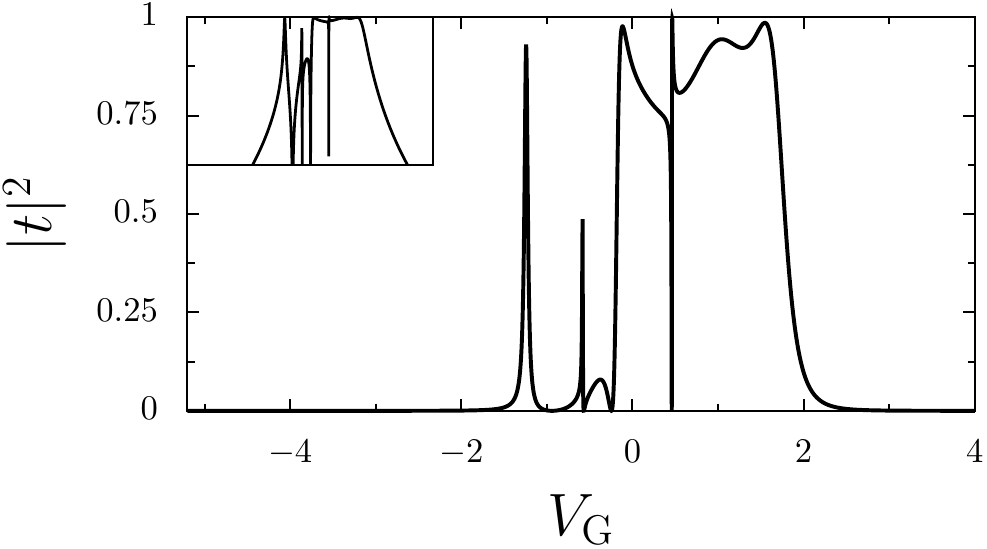}}
\centerline{\includegraphics[width=\figwidth]{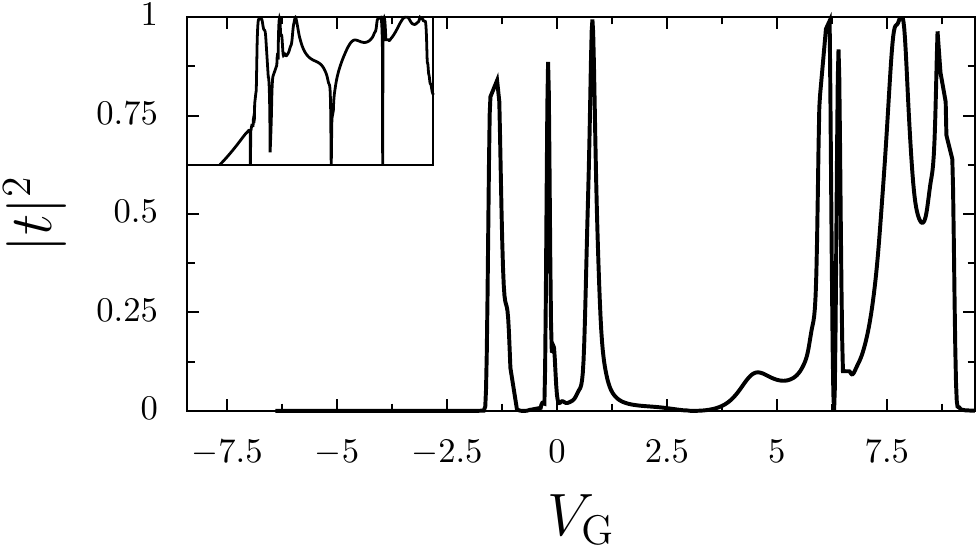}}
\caption{\label{fig:8site2} $|t|^2$ \textit{vs.} $V_\mathrm{G}$ of a lattice model for a quantum dot with 
$M=8$ sites and $U=0$ (4) in the top (bottom) panel.
$t_\mathrm{D}=0.5$ and $W=1$. The inset shows the same data on a logarithmic scale.}
\end{figure}
In general, the main effect of the interaction is to separate the resonance peaks and to make them narrower. 
In the more common scenario the number of transmission zeros is unchanged from the non-interacting to the 
interacting case. In some cases, the structure of the resonances may substantially change when interactions are 
switched on and the position of the zeros can change accordingly. Resonances that are very close and then 
indistinguishable in the non-interacting case, can often be resolved in the interacting case. A scenario we 
have observed in several examples is the disappearance of transmission zeros beyond a given interaction strength. 
This situation arises because the original UOR zeros occurring in the regime $\Gamma \gg \Delta$ disappear as 
interactions change the ratio between $\Gamma$ and $\Delta$. Pushing a transmission zero outside the interval 
where the resonances are (like the example shown in Fig.\ \ref{fig:4site2}) is another common outcome of 
including interactions. All cases analyzed resulted in the mesoscopic regime with a random alternation 
between resonances and transmission zeros and significantly less transmission zeros than resonances. In no instance 
did we find that interactions can induce a transition from ROR to UOR behavior as it was the case in the RM. 
Scaling up the system size or varying $t_\mathrm{D}$ and $U$ did not change these conclusions.

\begin{figure}
\centerline{\includegraphics[width=\figwidth]{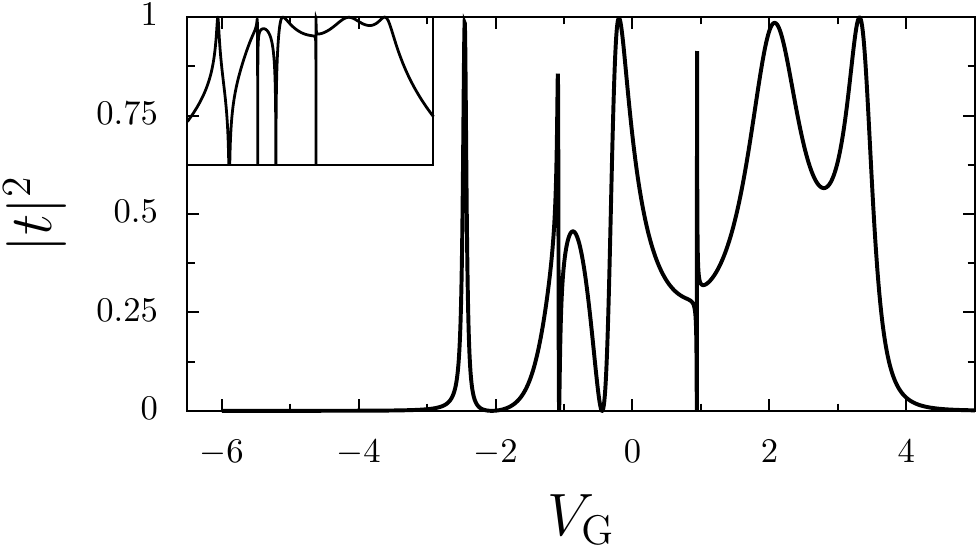}}
\centerline{\includegraphics[width=\figwidth]{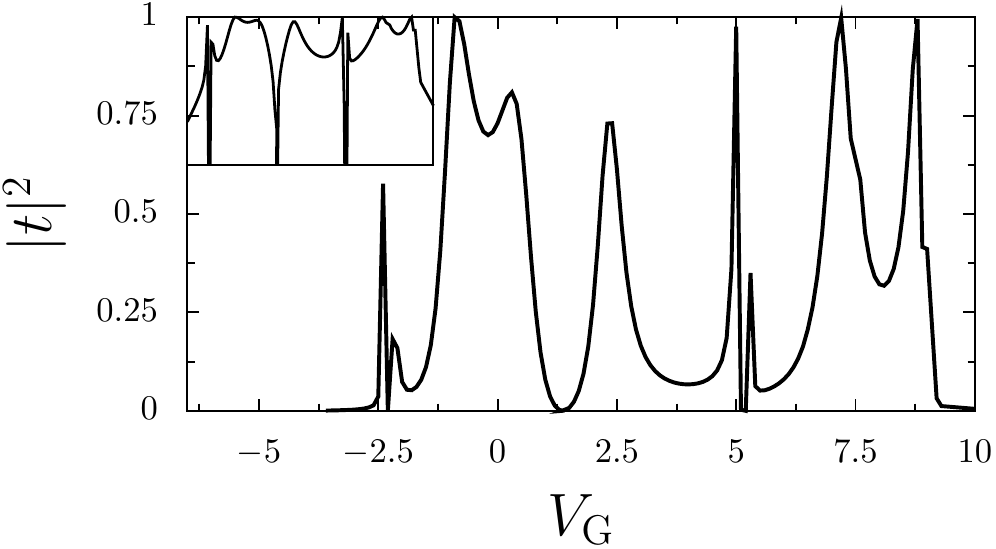}}
\caption{\label{fig:8site} $|t|^2$ \textit{vs.} $V_G$ for $U=0$ ($2$) in top (bottom) panel,
of a lattice model for a quantum dot with eight sites, $t_\mathrm{D}=1$, and $W=1$. 
The inset shows the same data on a logarithmic scale.}
\end{figure}
In Fig.\ \ref{fig:8site2} we show an example with $M=8$, $U=2$, and $t_\mathrm{D}=0.5$ exhibiting the most 
commonly encountered behavior where the number of zeros does not change from the non-interacting to the 
interacting case. Even if the structure of the resonances and the position of the zeros change dramatically 
from one case to the other, the total number of zeros remains constant. 

We finally illustrate in Fig.\ \ref{fig:8site} a scenario we often observed, with less zeros in the interacting 
case than in the non-interacting one in a quantum dot with $M=8$, $U=2$, $W=1$, and $t_\mathrm{D}=1$. The double 
zero in the middle of the curve (around $V_\mathrm{G}=1$) of the $U=0$ case corresponds to an UOR case that is 
transformed to a normal ROR zero for $U=2$. As resonant peaks become narrower and more separated in the presence 
of interactions UOR behavior is not favored. This is very different from the situation in the RM, and we do not 
observe here the population switching mechanism.

\section{Conclusions}
\label{Sec:conclusions}

In this work, we have presented numerical investigations of the transmission phase through confined, 
strongly correlated electron systems. Our goal was to determine how correlation effects influence 
the transmission phase, and whether they are able to explain the existing experimental observations of Ref.[4,6,9,10].
In order to compute the complex transmission amplitude of the scattering matrix 
corresponding to strongly correlated systems we have extended the embedding method previously used for 
conductance computations. For the simplicity of the numerical calculations we have used a model of spinless 
electrons. This is appropriate since spin effects are only expected to be important for extremely small dots, 
while our interest is to achieve the conditions of dots large enough to reach the transition from the mesoscopic 
to the universal regime. 

We have analyzed two different models including electron-electron interactions: the resonant level model and 
lattice models with nearest-neighbor interactions. The non-interacting limits of these models were also discussed 
in order to define different scenarios for the occurrence of transmission zeros and phase lapses. When the width 
of the resonances is smaller than the level spacing, the peaks are well resolved in energy. Two cases should then 
be considered: restricted off-resonance (ROR) behavior, in which the two nearest resonances determine the character 
of the transmission amplitude in-between; and unrestricted off-resonance (UOR) behavior, where the transmission 
amplitude between two adjacent resonances is significantly affected by other far-away resonances. In the ROR case 
we have one transmission zero or none, depending on the sign of $D_m$ [see Eq.\ (\ref{signrule})] being positive or 
negative, respectively. In the UOR case, and when the resonances are well separated, the number of zeros might be 
increased with respect to the ROR case by a multiple of two, which would not change the total phase shift 
accumulated in the interval between the resonances. Therefore, we have a phase shift of $\pi$ or $0$ between the 
resonances depending on $D_m$ being positive or negative, respectively. 

The situation changes when the resonances are strongly overlapping and it has no meaning to treat peaks and valleys 
separately in the energy dependence of the transmission amplitude. In these cases the transmission is typically 
given by the contributions from many resonances resulting in an UOR behavior with the possible appearance of zeros.
However, the phase accumulated in the region of overlapping peaks is not an integer multiple of $\pi$.
The transmission zeros cut the phase evolution leading to the incomplete filling of the dot and phase shifts 
smaller than $\pi$. This case is clearly not representative of the experimental situation, where well-resolved 
peaks alternate with conductance valleys.  

In the case of the resonant model we confirm previous results \cite{Karrasch07} about the possible increase in the 
number of zeros due to the presence of interactions in the regime of very wide resonances, depending on the choice 
of the coupling parameters. We relate this phenomenon to the UOR behavior studied in the non-interacting case. 
However, results for the RM do not fully reproduce key features of the experiment like the in-phase behavior of 
consecutive resonances. 

We treated lattice models of dots with up to $M=8$ sites. In most of the cases the number of transmission zeros was 
independent of the interaction strength, while in a minority of cases we have observed that interactions can induce 
ROR behavior from UOR behavior as resonances become more narrow and isolated, thus reducing the number of 
transmission zeros. Consistently with the experimental findings, we observe the mesoscopic behavior for the small 
size QD that we treat numerically. In addition, when decreasing the internal hopping amplitude to achieve smaller 
energy level separation with respect to the level couplings, no tendency towards universality was obtained, 
independently of the value of the interaction strength. The exploration of a large parameter space of level 
separations, coupling widths, and level spacings allowed to approach the conditions of not so small QD, where 
the universality was claimed to arise by the effect of electronic correlations. This is not the case. Only at large 
enough QD ($kL_\mathrm{S}\gg 1$), the one-particle wave-function correlations provoke the emergence of universality, 
but for those relatively large sizes the electronic correlations are no longer important.\cite{AlhassidRMP2000}

In the non-interacting case with chaotic underlying dynamics, the generic distribution of eigenstates and partial 
widths \cite{AlhassidRMP2000} favor the ROR behavior. Taking interactions at the CIM level may lead to the UOR 
behavior, but with well separated resonances. Therefore, in these cases the sign rule based on the one-particle 
wave-functions determines the phase behavior. The case of UOR with overlapping resonances is achieved by some 
tuning of the system parameters in the non-interacting and CIM cases.

Our main conclusion is that strong correlations cannot generically explain the experimentally observed universal
behavior of transmission phases in large dots, nor the crossover from mesoscopic behavior in few-electron dots to
universal behavior in many-electron dots.
This result is consistent with the observation\cite{Molina12} that the emergence of 
the universal behavior can be obtained taking into account one-particle wave-function correlations.

\acknowledgments

We acknowledge support from the
Spanish MICINN through project FIS2009-07277, the NSF under
grant No DMR-0706319, the ANR through grant ANR-08-BLAN-0030-02,
and the Swiss NCCR MANEP.

\appendix

\section{Embedding method for the transmission phase}
\label{Sec:embedding}

In this appendix we reformulate the embedding approach for the transmission phase put forward in 
Ref.\ \cite{Molina12a}, settling the notation and the basis of the numerical method used in 
Secs.\ \ref{Sec:mprlm} and \ref{Sec:mplm}. We also address the connection between scattering phase and 
induced charge, which are shown in App.\ \ref{Sec:appb} to provide a useful numerical test of the method 
beyond those used in Ref.\ \cite{Molina12a}.

The embedding method is a powerful technique to calculate the conductance through a strongly correlated 
nanosystem with or without disorder.\cite{molina03,molina04,MeSch03,vasseur06,vasseurthesis,molina05} 
The system of interest is connected to 
a one-dimensional lead that closes into itself resulting in a ring which is pierced by a magnetic field. The 
response of ground state properties to such a perturbation, like the persistent current or the phase sensitivity, 
allows to infer the conductance of the original system. While the one-dimensional setups with spinless electrons 
have been the most commonly used models, quasi-one dimensional leads and Hubbard chains have been recently 
considered,\cite{Freyn10} and nanosystems with non-trivial structure have also been studied.\cite{Moliner11}
The generalization of the embedding approach to the transmission phase lies on the same basis as 
the original method and provides a very useful tool.

The setup of the embedding method is given 
by a Hamiltonian as the one of Eq.\ \eqref{Eq:generalH}, where $H_\mathrm{D}$ stands for the Hamiltonian of the 
quantum dot depending on the model considered. $H_\mathrm{G}$ \eqref{Eq:Hg} allows for the application of a gate 
voltage, while the coupling term is given by \eqref{Eq:HClattice} or \eqref{Eq:HcoupRM}. The lead Hamiltonian 
needs to be modified with respect to \eqref{Eq:Hleads} in order to represent a ring pierced by a flux 
$\phi_\mathrm{e}$. It reads
\begin{equation}
H_\mathrm{L} = - \sum_{i=M+1}^L \left( c^{\dagger}_{i+1} c^{\phantom{\dagger}}_{i} + h.c. \right)
\end{equation} 
with the boundary condition $c_0=\exp(2\pi i\phi_\mathrm{e}/\phi_0)c_L$. In the inset of Fig.\ 
\ref{fig:ls10dotlhat5} we show the embedding setup for a linear QD (a chain) as used in App.\ \ref{Sec:appb} 
(notwithstanding a QD of arbitrary shape like those of Figs.\ \ref{Fig:LMsketch} and \ref{fig:RLMsketch} can 
be treated). Despite the similarity between the embedding setup and that of the AB interferometer, they are 
very different since the first is a closed system with fixed number of particles and the flux $\phi_\mathrm{e}$ 
is an auxiliary one, without physical reality.

Staying at first within a one-particle model (that is, without the interaction terms \eqref{Eq:HDrlint} 
or \eqref{Eq:HDlmint}) and with one-dimensional leads, the transport through the nanosystem is 
characterized by the scattering matrix  
\begin{equation}
S =\left(\begin{array}{cc} r & t' \\ t & r' \end{array}\right)
= e^{i \zeta} \left(\begin{array}{cc}
ie^{i\xi}\cos{\theta} & e^{i\eta}\sin{\theta} \\
e^{-i\eta}\sin{\theta} & ie^{-i\xi}\cos{\theta}
\end{array}\right)\, .
\label{eq:sm}
\end{equation}
We have chosen a generic parametrization of a $2\times 2$ unitary matrix. The transmission amplitude
for particles coming from the left of the QD and as defined in the introduction is given by
\begin{equation}
t =|t|e^{i\alpha} = e^{i (\zeta-\eta)}\sin{\theta}\, .
\label{eq:ta}
\end{equation}
It is related with the Green function by Eq.\ \ref{Eq:LeeFisher}. $t'$ is the transmission amplitude for 
particles impinging from the right and $r(r')$ is the reflection amplitude for particles coming from the 
left (right) of the QD. The angle $\theta$ and the scattering phase $\zeta$ are restricted to the interval
$[0,\pi)$, while the phases $\eta$ and $\xi$ are defined on $[0,2\pi)$. 

When the Hamiltonian of the dot exhibits time-reversal symmetry, one has $\eta=0$ or $\eta=\pi$. This will be
our case, since the artificial flux $\phi_\mathrm{e}$ used to drive the persistent current is seen by the ring, 
but not by the dot. When a control parameter is varied (like $V_\mathrm{G}$ or $\epsilon$), a jump of $\eta$ 
between its two possible values can only occur when the transmission amplitude vanishes ($\sin\theta=0$), in 
order to preserve the continuity of the scattering matrix. This observation is equivalent to that of Sec.\ 
\ref{Sec:Introduction} about the crossing of the origin of the complex plane by the transmission amplitude 
being typically associated with a jump of $\pi$ of its phase $\alpha$. 

Right-left symmetry would restrict $\xi$ to $0$ or $\pi$. Similarly as in the case of time-reversal symmetry, 
jumps of $\pi$ in $\xi$ are only allowed when $\cos\theta=0$, that is, when the reflection amplitude vanishes. 
However, throughout this work, we consider generic QDs with arbitrary $\xi$.

When considering the phase evolution as a function of an external parameter, it is often convenient to work with 
the accumulated phase $\alpha_\mathrm{c}$, whose range of definition is not restricted to the interval $[0,2\pi)$.

Embedding the scatterer in a ring of length $L_\mathrm{L}$ pierced by a dimensionless flux 
$\Phi_\mathrm{e}=2\pi\phi_\mathrm{e}/\phi_0$ leads to the following quantization condition for the
one-particle states of the composed system\cite{molina04}
\begin{equation}
\cos(\Phi_\mathrm{e}-\eta)=\frac{1}{\sin{\theta}}
\cos\big( kL_\mathrm{L}+\zeta\big) \ .
\label{eq:quantcont2}
\end{equation}
Since the scattering phase $\zeta$ belongs to the interval $[0,\pi)$, there are two branches for the solutions 
(in $k$) of \eqref{eq:quantcont2} corresponding to the two possible values of $\eta$ (0 and $\pi$). On the 
other hand, the transmission phase $\alpha$ is defined in $[0,2\pi)$, which allows to write 
\eqref{eq:quantcont2} in the more compact way 
\begin{equation}
\cos(\Phi_\mathrm{e})=\frac{1}{\sin{\theta}}
\cos\big( kL_\mathrm{L}+\delta\alpha \big) \ .
\label{eq:quantcont3}
\end{equation}
We note $L_\mathrm{S}$ the length of the scatterer between the leads, $L=L_\mathrm{S}+L_\mathrm{L}$ the total 
length of the ring, and the phase shift $\delta\alpha=\alpha-kL_\mathrm{S}$. We express $L$ and $k$ in Anderson 
units (that is, in terms of the lattice spacing). Knowing the dispersion relation in the leads $\epsilon(k)$, the 
sum over the energetically lowest one-body energies $\epsilon(k_n)$ allows to obtain the ground-state energy 
$E(N)$ of the whole system containing $N$ particles. For an odd number of particles $N=2n_\mathrm{F}+1$, the 
lowest order terms in a $1/L$ expansion read \cite{molina04}
\begin{equation}\label{eq:energy}
E(N)=E^{(0)}(N)-\frac{2}{L}\sum_{n=1}^{n_\mathrm{F}}
\left.\frac{\mathrm{d}\epsilon}{\mathrm{d}k}\right|_{k=k_n^{(0)}}
\delta\alpha(k_n^{(0)})+O[1/L]
\end{equation} 
with the $k$-values in a clean ring $k_n^{(0)}=2\pi n/L$. The ground state energy of a clean ring (with a 
scatterer having perfect transmission $|t|=1$ and $\delta\alpha=0$) is given by
\begin{equation}\label{eq:energy0}
E^{(0)}(N)=\epsilon(0)+2\sum_{n=1}^{n_\mathrm{F}}\epsilon(k_n^{(0)})\, .
\end{equation}
The embedding method allows to obtain the conductance of the scatterer from the flux dependence of $E(N)$, which 
only appears in second order in $1/L$. By changing the total number of particles, we have 
access, with the help of \eqref{eq:energy} and \eqref{eq:energy0}, to the scattering phase shift at the Fermi 
level ($k_\mathrm{F}=\lim_{L\to\infty} k_{n_\mathrm{F}}^{(0)}$) of the lead
\begin{equation}\label{eq:phase-general}
\delta\alpha(k_\mathrm{F})=-\lim_{L\to\infty} \frac{L}{2}
\left(\frac{E(N)-E(N-2)-2\epsilon(k_{n_\mathrm{F}}^{(0)})}
{\left.\mathrm{d}\epsilon/\mathrm{d}k\right|_{k=k_{n_\mathrm{F}}^{(0)}}}\right) \, .
\end{equation}
In a chain with $N=L/2+1$ (and $L$ multiple of 4) we are effectively at half filling, 
$k_{n_\mathrm{F}}^{(0)}= \pi/2$ such that Eq.\ \ref{eq:phase-general} reduces to 
\begin{equation}\label{eq:phase}
\delta\alpha(k_\mathrm{F})= - \lim_{L\to\infty} \frac{L}{4} \left[E(L/2+1)-E(L/2-1)\right]\, .
\end{equation}
The previous derivation is based on a single-particle approach. As in the case of the embedding method for the 
conductance, the passage to the many-body system is justified by the fact that the scattering properties of a 
many-particle scatterer can be represented by an effective one-particle scattering matrix. For instance, it has 
been verified \cite{molina04,Freyn10} that the flux-dependence of the many-particle ground state is, in the large 
$L_\mathrm{L}$ limit, reproduced by the total energy obtained from effective single-particle states (as done in 
the previous derivation).

In the many-body case, the scattering matrix \eqref{eq:sm} is therefore understood as an effective one, where each of its 
entities depends on the interaction strength $U$. This identification is made in the embedding method for 
the transmission phase as well as for the conductance. As a logical consequence, the results for the two quantities 
have to be consistent. Indeed, we obtain jumps in the transmission phase as a function of the parameters precisely at the 
positions where transmission zeros occur.
When using Eqs.\ \ref{eq:phase-general} or \ref{eq:phase} in 
order to obtain the transmission phase, the limiting procedure is implemented by extrapolating towards large 
values of $L_\mathrm{L}$. This procedure is numerically demanding, and constitutes the bottleneck of the 
embedding method.\cite{molina03,molina04,Molina12a,molina05,vasseur06,vasseurthesis,Freyn10}

The eigenphases of the scattering matrix are, for $\cos\theta\cos\xi > 0$, 
$\varphi_1=\zeta+\mathrm{Arcsin}(\cos\theta\cos\xi)$ and $\varphi_2=\zeta+\pi-\mathrm{Arcsin}(\cos\theta\cos\xi)$,
and the corresponding Wigner time is
\begin{equation}\label{eq:wt}
\tau(\epsilon)=\frac{\hbar}{2}\sum_{q=1}^{2} \frac{\mathrm{d} \varphi_{q}}{\mathrm{d} \epsilon} =
\hbar \ \frac{\mathrm{d}\zeta}{\mathrm{d}\epsilon} \ .  
\end{equation}
The one-particle density of states in the scattering region is related to the Wigner time as \cite{DoSm,JP}
\begin{equation}\label{eq:doswt}
d(\epsilon)=\frac{1}{\pi \hbar} \langle \tau(\epsilon) \rangle \, ,
\end{equation}
where the brackets stand for a spectral average over many eigenstates.

Integrating over an energy interval we recover the Friedel sum rule \cite{friedel52,langer61}
\begin{equation}\label{eq:friedel}
\Delta \zeta=\pi \Delta N_\mathrm{S}
\end{equation}  
as a relationship between the number of particles $N_\mathrm{S}$ added to the scattering region and the 
corresponding change in the scattering phase.

The fact that the Friedel sum rule in its form \eqref{eq:friedel} applies to the scattering phase $\zeta$ has 
been emphasized in the literature \cite{Lee99,levy00,taniguchi99}. The lapses of $\pi$ in the transmission 
phase $\alpha$ at the zeros of $t$ are not related with a special behavior of the density of states. However, 
the integration leading to \eqref{eq:friedel} involves a large energy interval (on the scale of the level spacing) 
where many lapses appear. 
Since the origin of the complex plane is crossed in many different directions, 
the effect of the lapses tends to average out, and we can write the accumulated phase in the interval as also 
given by the Friedel sum rule
\begin{equation}\label{eq:friedel2}
\Delta \alpha_\mathrm{c}=\pi \Delta N_\mathrm{S} \, .
\end{equation}  
Such an average behavior has been discussed in Ref.\ [\onlinecite{Molina12}] where the ambiguity between the lapses 
of $\pi$ and $-\pi$ was proposed to be lifted by applying a small magnetic field. Then, the origin in 
the complex $t$-plane can be avoided and well defined phase lapses with a phase change close to $\pi$ or $-\pi$ obtained. The 
scattering phase can then be obtained from the accumulated phase by taking 
$[\alpha_\mathrm{c}(B=0^+)+\alpha_\mathrm{c}(B=0^-)]/2$.

The existence of phase jumps in the $V_\mathrm{G}$-dependence of $\alpha$ can also be obtained from the standard 
embedding method applied to the conductance by locating the zeros of the transmission, as we do in Sec.\ 
\ref{Sec:mplm}. The quantization condition \eqref{eq:quantcont2} leads to a phase sensitivity\cite{molina04}
\begin{equation}\label{eq:phsen}
\begin{aligned}
\Delta E &=E(N,\Phi_\mathrm{e}=\pi)-E(N,\Phi_\mathrm{e}=0)\\
&=\frac{\hbar v_\mathrm{F}}{L} \left[\frac{\pi}{2}-\mathrm{Arccos}(\sin{\theta}\cos{\eta})\right] \, ,
\end{aligned}
\end{equation}
and therefore
\begin{equation}\label{eq:sinthsinphi}
|t|= \sin{\left(\Delta E \frac{L}{\hbar v_\mathrm{F}}\right)}\cos{\eta}=
     \sin{\left(\frac{\pi}{2} \left|\frac{\Delta E}{\Delta E^{(0)}}\right|\right)} \, ,
\end{equation}
where $\Delta E^{(0)}$ is the phase sensitivity of a perfectly transmitting scatterer. Since we are working with 
time-symmetric dots, $\eta$ can only take the values $0$ or $\pi$. The switches between these two branches may 
only occur when $t=0$. Therefore, the sign changes of $\Delta E$ are associated with the $\pi$ lapses in $\eta$ 
and $\alpha$. One should also notice that there can also be transmission zeros without phase lapses in cases when 
there is a zero of $\Delta E$ without a sign change. Indeed this possibility can occur for particular values of 
the parameters, as discussed in Sec.\ \ref{Sec:1plattice}.

For a strictly one-dimensional system, the sign of $\Delta E$ is fixed by Leggett's theorem \cite{leggett,waintal08}. 
For an odd number of particles we are in the branch of $\eta=0$, the transmission amplitude $t$ never vanishes, and 
there cannot be branch-switches or phase lapses. For a quasi-one dimensional scatterer this is no longer true, and 
we expect to have parameter values where $t$ vanishes and branch-switches appear. 

\section{Transmission phase of a one-dimensional many-body scatterer}
\label{Sec:appb}

The applicability of the embedding method to obtain the transmission phase of a many-body scatterer can be 
conveniently tested in the one-dimensional case. As discussed at the end of \ref{Sec:embedding}, one-dimensional 
systems are constrained to the branch $\eta=0$, thus $\alpha=\zeta$, and there are no transmission zeros. On the 
other hand, comparing the numerical results to the prediction from the Friedel sum rule constitutes a valuable 
test of the method. 

An interacting one-dimensional chain is a particularly simple example of a many-particle lattice model where the 
dot Hamiltonian (Eq.\ \ref{Eq:HDlattice}) only has $M=L_\mathrm{S}$ ordered sites. For simplicity, we work in this 
section in the non-disordered case $\epsilon_i=0$ and we take $t_\mathrm{D}=1$. In the absence of a gate voltage 
$(V_\mathrm{G}=0)$, we work at half filling, $E(L/2+1)=E(L/2-1)$, for periodic boundary conditions in the ring and 
thus $\delta\alpha=0$, independent of the interaction strength. The number of electrons in the scattering region 
\begin{equation}
N_\mathrm{S}=\sum_{i=1}^{L_\mathrm{S}}\langle n_{i}\rangle
\end{equation}
is equal to $L_\mathrm{S}/2$. This is consistent with the findings of Ref.\ [\onlinecite{molina05}], where 
Fabry-Perot like oscillations of the transmission through two interacting regions in series were studied. 

\begin{figure}
\centerline{\includegraphics[width=\linewidth]{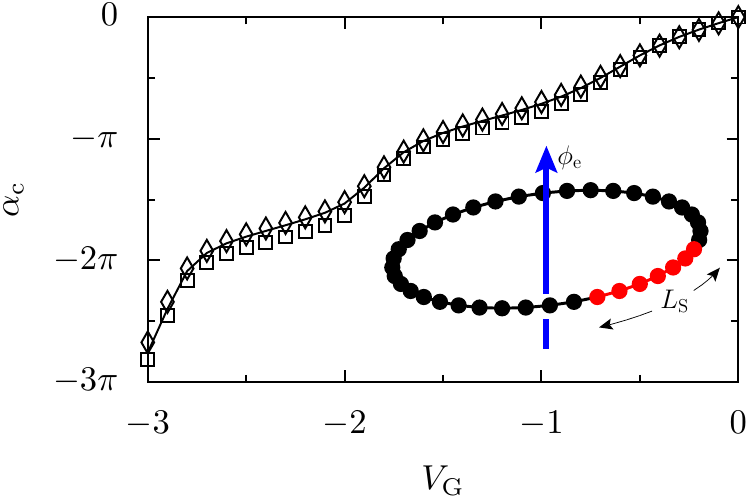}}
\caption{(Color online) Accumulated transmission phase shift for a many-body scatterer of length $L_\mathrm{S}=10$ 
with interaction strength $U=2$ (solid line), as a function of the gate voltage $V_\mathrm{G}$. 
The squares are the prediction of the Friedel sum rule based on the numerically obtained charge densities 
on the $L_\mathrm{S}$ interacting sides. The diamonds result when the densities on the 
$L_\mathrm{S}$ sites of the scatterer plus that of 10 additional sites surrounding the interacting region.
The inset shows a sketch of the embedding setup for a simple one-dimensional interacting scatterer (in red).}
\label{fig:ls10dotlhat5}
\end{figure}
In contrast, once an additional gate voltage is applied (Eq.\ \ref{Eq:Hg}), particle-hole symmetry is broken and 
$N_\mathrm{S}$ will differ from $L_\mathrm{S}/2$. The phase shift is thus expected to be nonzero. For such a setup, 
the embedding method has been used to show that Coulomb-blockade like oscillations of the conductance as a function 
of $V_\mathrm{G}$ appear in the presence of interactions \cite{vasseur06} even in the well coupled case. The charge 
in the dot region decreases in steps once $V_\mathrm{G}$ is increased and depletes the interacting region. We extend 
now these  DMRG \cite{DMRG_book,schmitteckert_thesis} based calculations to compute the ground state density of rings 
embedding such a many-body scatterer, as well as the transmission phase resulting from \eqref{eq:phase}. 
We choose a chain with $L_\mathrm{S}=10 $ sites and an interaction strength $U=2$.

Data for ring sizes up to $L=120$ were used in the extrapolation towards infinite size. The results for the 
transmission phase are presented in Fig.\ \ref{fig:ls10dotlhat5} (solid line) together with 
$\pi N_\mathrm{S} - \pi  L_\mathrm{S}/2$ (squares).

Both quantities should be equal according to the Friedel sum rule since $\delta\alpha=\alpha-\pi L_\mathrm{S}/2$ 
for half filling. The results have very similar behavior but small quantitative differences appear. 

The difference disappears when the density modifications outside the $L_\mathrm{S}$ sites of the scattering 
region are included in the calculation of $N_\mathrm{S}$. Taking into account the density changes in the 
interacting region plus that on 5 additional sites on either side of the scatterer (diamonds), the values of 
$\pi N_\mathrm{S} - \pi  L_\mathrm{S}/2$ are in quantitative agreement with those of $\delta\alpha$.

From this numerical example we have learned how precise the embedding method for scattering phases is, and we 
have checked that in order to comply with the Friedel sum rule, all the charge displacement in the neighborhood of 
the scatterer has to be accounted for. The effect of the charge build-up in the leads of the AB interferometer was 
invoked in Ref.\ [\onlinecite{levy95}] as an important ingredient in order to address the physics of the 
experimentally observed in-phase behavior of consecutive resonances. This kind of charge displacement and 
screening effects  might be responsible for the phase increase 
smaller than $\pi$ at certain resonances which is observed in the experimental data.\cite{Schuster97} 

The embedding method is particularly efficient when dealing with one-dimensional leads \cite{molina04,Freyn10},
but the scatterer might have any topology or dimensionality. Upon this fact is based our numerical work 
of Secs.\ \ref{Sec:mprlm} and \ref{Sec:mplm} where the quasi-one dimensional systems that allow transmission 
zeros are thoroughly studied.


\begin{thebibliography}{99}
\bibitem{numericaldifficulty}
For the larger models, we had to calculate ring sizes up to $L=136$ keeping 1600 
states in the DMRG algorithm to obtain good convergence of the embedding method. The full calculation 
of one of the transmission curves could amount to approximately 20000 CPU hours in an Intel Xeon processor. 
We were able to calculate a few different realizations of disorder for each set of parameters $U=2,4,8$, 
$t_D=0.25,0.5,1$, and $M=4,6,8$. Calculations of $|t|$ converge faster than those of the full 
complex $t$.
\end{thebibliography}
\end{document}